\documentclass[10pt,journal,compsoc]{IEEEtran}

\usepackage{epsfig}
\usepackage{subfigure}
\usepackage{epstopdf} 
\usepackage{amssymb}
\usepackage{lscape,graphicx}
\usepackage{times}
\usepackage{url}
\usepackage{amsmath}
\usepackage{enumerate}
\usepackage{multirow}
\usepackage[normalem] {ulem} 
\usepackage{graphics}
\usepackage[table]{xcolor}
\usepackage[linesnumbered, algoruled, vlined]{algorithm2e}
\usepackage{lineno,hyperref}

\SetKwInOut{Input}{input}
\SetKwInOut{Output}{output}
\SetNlSty{texttt}{\color{red}}{}
\SetFuncSty{textit}
\SetAlFnt{\footnotesize}
\SetAlgoNlRelativeSize{-2}
\SetKwComment{tcp}{\color{gray}// }{}

\def\upto{\mathbin{\mathstrut{\ldotp\ldotp}}}

\newcommand{\nm}[1]{#1}

\begin{document}
%
\title{Performance Modeling of Microservice Platforms}
%
%


\author{Hamzeh~Khazaei,~\IEEEmembership{Member,~IEEE,}
        Nima~Mahmoudi,~\IEEEmembership{Student Member,~IEEE,}
        Cornel~Barna,~\IEEEmembership{Member,~IEEE,}
        and~Marin~Litoiu,~\IEEEmembership{Senior~Member,~IEEE}
\IEEEcompsocitemizethanks{\IEEEcompsocthanksitem H. Khazaei is with the Department of Electrical Engineering and Computer Science,
York University, Toronto, Ontario, Canada.\protect\\
E-mail: hkh@yorku.ca
\IEEEcompsocthanksitem N. Mahmoudi is with the Department
of Electrical and Computer Engineering, University of Alberta, Edmonton,
Alberta, Canada.\protect\\
E-mail: nmahmoud@ualberta.ca
\IEEEcompsocthanksitem C. Barna is with Seneca Collage.
\IEEEcompsocthanksitem M. Litoiu is with the Department of Electrical Engineering and Computer Science,
York University, Toronto, Ontario, Canada.\protect\\
E-mail: mlitoiu@yorku.ca}
\thanks{Manuscript received April XX, 2020; revised August XX, 2020.}}
       

\markboth{IEEE Transactions on Cloud Computing,~Vol.~xx, No.~x, August~2020}%
{Khazaei \MakeLowercase{\textit{et al.}}: Performance Modeling of Microservice Platforms Considering the Dynamics of the underlying Cloud Infrastructure}

\IEEEcompsoctitleabstractindextext{%
\begin{abstract}
Microservice architecture has transformed the way developers are building and deploying applications in the nowadays
cloud computing centers. This new approach provides increased scalability, flexibility, 
manageability, and performance while reducing the complexity of the whole software development life cycle. 
The increase in cloud resource utilization also benefits microservice providers. Various microservice platforms
have emerged to facilitate the DevOps of containerized services by enabling continuous integration and delivery. 
Microservice platforms deploy application containers on virtual or physical machines provided by public/private cloud  
infrastructures in a seamless manner. In this paper, we study and evaluate the provisioning performance of microservice 
platforms by incorporating the details of all layers (i.e., both micro and macro layers) in the modelling process.
To this end, we first build a microservice 
platform on top of Amazon EC2 cloud and then leverage it to develop a comprehensive performance model to perform 
what-if analysis and capacity planning for microservice platforms at scale. In other words, the proposed performance 
model provides a systematic approach to measure the elasticity of the microservice platform by analyzing the provisioning 
performance at both the microservice platform and the back-end macroservice infrastructures.

\end{abstract}

}

\maketitle

\IEEEdisplaynotcompsoctitleabstractindextext

%
\IEEEpeerreviewmaketitle

\section{Introduction} \label{intro}
\IEEEPARstart{V}{irtual}
Machines (VM) are a widely used building block of workload management and deployment. They are heavily
used in both traditional data center environments and clouds. A hypervisor is usually used to manage all virtual machines on a physical machine. This virtualization technology is quite mature now and can provide good performance and security isolation among 
VM instances. The isolation among VMs is strong, and an individual VM has no awareness of other VMs running on the same physical machine (PM). However, for applications that require higher flexibility at runtime  and less isolation, 
hypervisor-based virtualization might not satisfy the entire set of quality of service (QoS) requirements.

Recently there has been an increasing interest in container technology, which provides a more lightweight
mechanism by way of operating system level virtualization compared to VMs~\cite{khan2017orch}. 
A \textit{container} runs on a kernel with similar performance isolation and allocation characteristics as VMs but without the expensive VM runtime management overhead \cite{soltesz2007container}. Containerization of applications, which is 
the deployment of application or its components in containers, has become popular in the cloud service 
industry~\cite{bernstein2014containers,google-cont, celesti2016exploring}. 
For example, Google, Amazon, eBay and Netflix are providing 
many of their popular services through a container-based cloud \cite{burns2016design}. 

A popular containerization engine is \textit{Docker}~\cite{merkel2014docker}, which 
wraps an application in a complete filesystem that contains everything required to run: 
code, runtime, system tools, and system libraries \textendash{} anything that can be installed on a VM. This guarantees that 
the application will always run the same, regardless of its runtime environment~\cite{merkel2014docker}. 
Container-based services are popularly known as \emph{Microservices} (while VM based services are referred to 
as \emph{Macroservices}) and are being leveraged by many service providers for a number of 
reasons: (a) reduced complexity for tiny services; (b) easy and fast scalability and deployment 
of application; (c)  improvement of flexibility by using different tools and frameworks; (d) enhanced reliability for the system~\cite{burns2016design}. 
Containers have empowered the usage of microservices architectures by being lightweight, providing 
fast start-up times, and having low overhead. Containers can be used to develop  \textit{monolithic 
architectures} where the whole system runs inside a single container 
or \textit{clustered architectures}, where a combination of containers is used~\cite{khazaei2016efficiency}. 

A flexible computing model combines Infrastructure-as-a-Service (IaaS) structure with container-based
Platform-as-a-Service (PaaS). Leveraging both containers and VMs brings the best of both technologies for all stakeholders: the strong isolation of VM for the security and privacy purposes, and the flexibility of containers to improve performance and 
management. Platforms such as Tutum \cite{tutum}, Kubernetes \cite{kubernetes}, Nirmata \cite{nirmata} 
and others~\cite{hook, vamp} offer services for managing microservice environments made of containers while relying on IaaS 
public/private clouds as the backend resource providers. Application service providers who plan to migrate to microservice
platforms need to know how their QoS would be at scale. Service availability and service response 
time are among the top two key QoS metrics~\cite{hamzeh-tpds}. Quantifying and characterizing such quality measures require 
accurate modelling and coverage of  large parameter space while using  tractable and solvable models in a timely manner to 
assist with runtime decisions \cite{longo:2011}. Scalability is a key factor in the realization of availability and responsiveness of a 
distributed application in times of workload fluctuation and component failure. In other words, the method and delay at which an application
can provision and deprovision resources determine the service response time and availability.  

In this work, we build a microservice platform using the most widespread technology, i.e. Docker~\cite{merkel2014docker}, and then 
leverage this platform to design a tuneable analytical performance model that provides what-if analysis and capacity planning 
at scale. Both microservice platform providers and microservice application owners may leverage the performance
model to measure the quality of their elasticity in terms of provisioning and deprovisioning resources. The proposed performance 
model is comprehensive as it models both microservice as well as the macroservice layers through separate
but interactive sub-models. The performance model supports a high degree of virtualization at both layers 
(i.e., multiple VMs on a PM and multiple containers on a VM) that reflects the real use case scenarios in today's microservice platforms. It is worth mentioning that in this paper the focus is on the provisioning performance of the
microservice platform and not the performance of deployed applications.   
A preliminary version of this work appeared as a conference paper~\cite{khazaei2016efficiency}, 
which only models the microservice layer and treats the back-end IaaS cloud as a black box. In this work, however, we propose a comprehensive 
fine granular performance model that captures the details of both layers and their interactions; we discuss the scalability and tractability 
and also provide an algorithm that presents the internal logic of the performance model.

The rest of the paper is organized as follows. Section \ref{vt} describes the virtualization
techniques and the new trend of emerging microservices. Section \ref{sd} describes the system 
that we are going to model in this work. Section \ref{lam} introduces the stochastic sub-models 
and their interactions. Section \ref{env} presents the experiments and numerical results obtained from the 
analytical model. In section \ref{rw}, we survey related work in cloud performance analysis, 
and finally, section \ref{con} summarizes our findings and concludes the paper.

\section{Virtualization and Microservices} \label{vt}
The concept of virtualization was created with the main goal of increasing the efficient use of computing resources. 
With the advent of cloud computing, virtualization has received even more attention for managing cloud data centers. 
Cloud providers essentially employ two main components,  cloud management systems and hypervisors,  to manage 
their physical stack of resources and make it accessible to a large number of users. Hypervisors provide a well 
defined logical view of the physical resources of a single physical machine (PM). Cloud management systems, on the other hand, 
provide a single interface of management for both service providers and cloud users of the whole data center. This way, 
cloud resources are provided to users as virtual machines (VMs),  highly isolated environments with a full operating system. 
The most popular hypervisors are VMware \cite{vmware}, KVM \cite{kvm}, Xen \cite{xen} and Hyper-V \cite{hyper-v}; 
and popular open-source cloud management systems include OpenStack \cite{open-stack}, CloudStack \cite{cloud-stack}, 
and Eucalyptus \cite{eucalyputs}.
Hypervisor-based virtualization provides the highest isolated virtual environment.
However, the cost of virtualization overhead is high, as each VM has to run its own kernel as the Guest OS. Moreover, 
VM resources are mainly underutilized, as each VM usually hosts one application \cite{felter2014updated}. VM virtualization 
limitations led to the development of Linux containers  wherein only resources required by applications will be used
while avoiding the overhead of redundant virtualized operating systems, as is the case in VMs.

A Linux container takes a different approach than a hypervisor and could be used as an alternative to or a complement for hypervisor-based virtualization. Containerization is an approach where one can run many processes in an isolated fashion. It uses only one kernel (i.e., OS) to create multiple isolated environments. Containers are very lightweight because they do not virtualize the hardware; instead, all containers on the physical host use the single host kernel efficiently via process isolation. Containers are highly portable, and applications can be bundled into a single unit and  deployed in various environments without making any changes to the container. Because of the standard container format, developers only have to worry about the applications running inside the container and system administrators will take care of deploying the container onto the servers. This well-segregated container management leads to faster application delivery. Moreover, building new containers is fast because containers are very lightweight, and it takes seconds to build a new container. This, in turn, reduces the time for DevOps, including development, test, deployment and runtime operations \cite{joy2015performance}. All in all, containers not only lead to significant improvements in the software development life cycle in the cloud, but also provide a better QoS
compared to non-containerized cloud applications. Several management tools are available for Linux containers, including LXC \cite{lxc}, lmctfy \cite{lmctfy}, Warden \cite{warden}, and Docker \cite{merkel2014docker}.

Recently, a pattern has been adopted by many software-as-a-service providers in which both VMs and containers are leveraged to provide so-called \textit{microservices}. Microservices is an approach that allows more complex applications to be configured from basic building blocks, where each building block is deployed in a container, and the constituent containers are linked together to form the cohesive application. The application's functionality can then be scaled by deploying more containers of the appropriate building blocks rather than entire new instances of the full application. \textit{Microservice platforms} (MSP) such as Nirmata \cite{nirmata}, Docker Cloud \cite{tutum} and Google Kubernetes \cite{kubernetes} facilitate the management of such service paradigms. MSPs are automating deployment, scaling, and operations of application containers across clusters of physical machines in the cloud. MSPs enable Software-as-Service (SaaS) providers to quickly and efficiently respond to customer demand by scaling the applications on-demand, seamlessly rolling out new features, and optimizing hardware
usage by using only the resources that are needed. 
\begin{figure}[!t]
\begin{center}
\includegraphics[width=0.95\columnwidth]{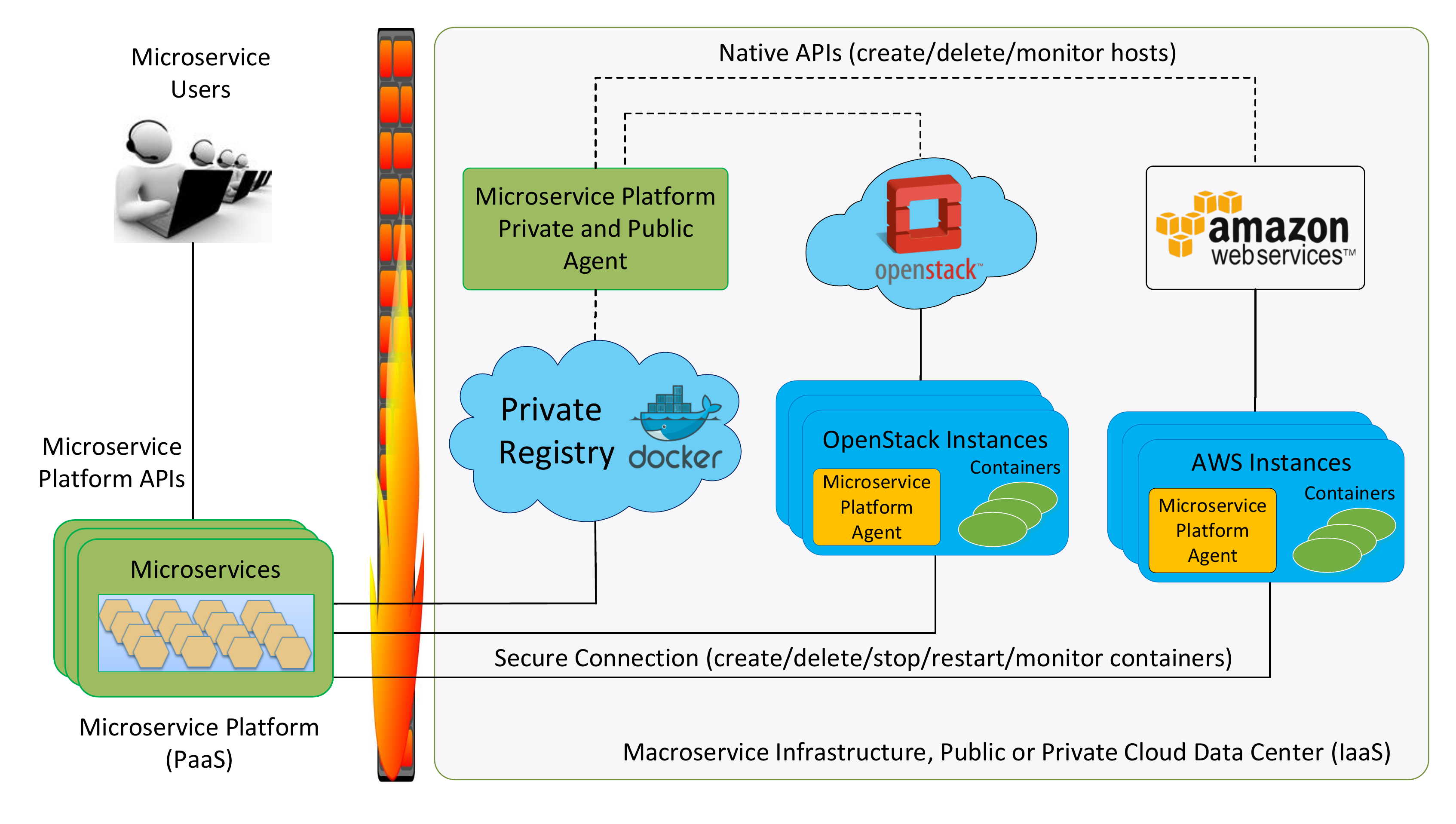}
\end{center}
\caption{Conceptual Model: including both the microservice platform and macroservice 
infrastructure (i.e., backend public/private cloud)~\cite{khazaei2016efficiency}.}
\label{fig:micro-concept-model}
\end{figure}
Fig. \ref{fig:micro-concept-model} depicts the high-level architecture of MSPs and the way they leverage the backend public or private cloud (i.e., infrastructure-as-a-service clouds). 


\begin{figure*}[!t]
\begin{center}
\includegraphics[width=0.9\textwidth]{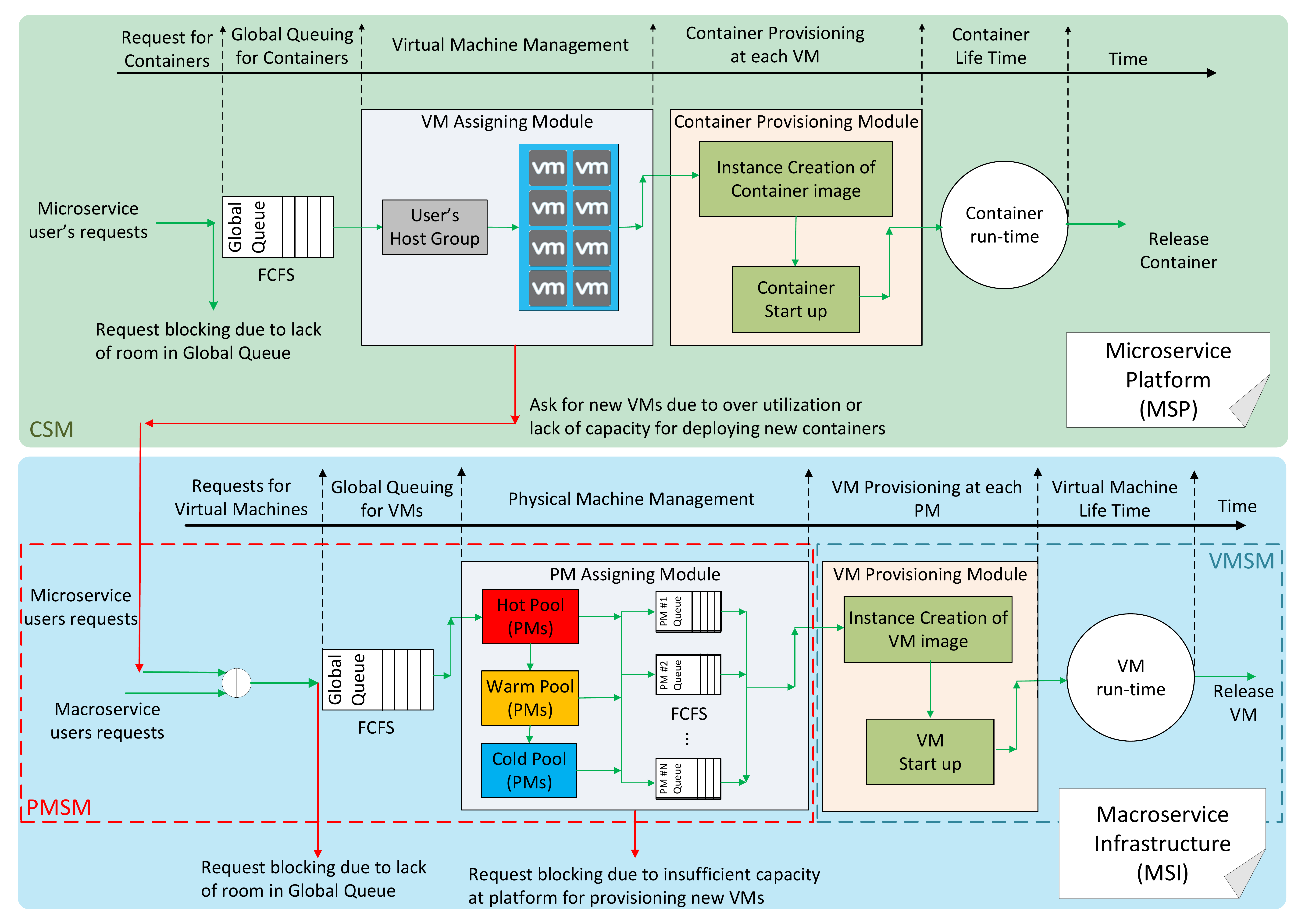}
\end{center}
\caption{The master performance model, derived form conceptual model in 
Fig. \ref{fig:micro-concept-model}. The three sub-systems reflected in the
performance model by each of the sub-models (CSM, PMSM, and VMSM) are annotated
in the figure.}
\label{fig:si-rt}
\end{figure*}

\section{System Description} \label{sd}
In this section, we describe the system under modelling with respect to Fig. \ref{fig:si-rt} 
that has been derived from the conceptual model shown in Fig. \ref{fig:micro-concept-model}. 
First, we describe the microservice platform (MSP) in which users request containers. In MSP, 
a request may originate from two sources: 1) direct requests from users that want to deploy 
a new application or service; and 2) runtime requests either directly from users
or from applications (e.g., consider adaptive applications) by which applications adapt to the 
runtime conditions, e.g., scaling up the number of containers to cope with a traffic peak.
In the proposed performance model, this platform is modelled by the container
sub-model (CSM).

Now consider a user, e.g., John, that wants to provide a new service for his customers. 
He would log into his MSP and create a Host Group with 2 initial VMs for the new services, setting the max size of host group to 5 VMs and each VM to run up to 4 containers.
In other words, his host group can scale up to 5 VMs that can accommodate up to 20 containers in total.  
He requests 6 containers for the initial setup and deploys 3 containers on each VM. 
Now, John has 2 active VMs, each running 3 containers. His application is adaptive so 
that in case of high traffic, it will scale up by adding one container at a time. 
Consider a situation in which John's application adds one more container to the application 
due to high traffic. At this time, one VM is working with full capacity (i.e., running 4 
containers), and another one only has capacity for one more container. Since John has configured 
his cluster application to grow up to 5 VMs, the MSP requests a new VM from the macroservice provider 
(i.e., IaaS) due to high utilization in the host group. Thus, John's 
host group now has 3 active VMs and can accommodate 5 more containers. Now consider 
the reverse scenario in which low traffic is entering the system and the application releases a 
few of containers after some time. If the host group utilization 
reaches a predefined low threshold, then the MSP would release one VM. We capture all these 
events for every single user in a Continuous-Time Markov Chain (CTMC) that will be described in section
\ref{mpm}. 

The steps incurred in servicing a request in MSP are shown in the upper part of Fig. \ref{fig:si-rt}. 
User requests for containers are submitted to a global finite queue and then processed 
on a first come, first served (FCFS) basis. A request that finds the queue full will 
be rejected immediately. Also, the request will be rejected if the application has already 
reached its capacity (for example, in John's scenario, if the host group already has 5 running VMs,
each of which is running 4 containers). Once the  request is admitted to the 
queue, it must wait until the VM Assigning Module (VMAM) processes it. VMAM finds the 
most appropriate VM in the user's Host Group and then sends the request to that VM so 
that the Container Provisioning Module (CPM) initiates and deploys the container (second delay). 
When a request is processed in CPM, a pre-built or customized container image is used to 
create a container instance. These images can be loaded from a public repository like Docker Hub 
\cite{merkel2014docker} or private repositories. 

In the macroservice infrastructure (lower part in Fig. \ref{fig:si-rt}), when a request is processed, 
a pre-built or customized disk image is used to 
create a VM instance \cite{hamzeh-tpds-fine}. In this work, we assume that pre-built 
images fulfill all user requirements and PMs and VMs are homogeneous. Each PM has 
a Virtual Machine Monitor (VMM), aka hypervisor, that configures and deploys VMs on a PM. 
In this work, we allow users to submit a request for a single VM at a time. 
Two types of requests are arriving at the Macroservice Infrastructure: 
1) requests that are originating from the cloud users, which are referred to as Macroservice users in
Fig. \ref{fig:si-rt}; and 2) requests from MSPs by which users/applications are asking for VMs to
deploy/scale their containers. Self-adaptive applications in MSP might autonomously and directly ask for VMs from 
the backend cloud. Since the number of users/applications in MSP is relatively high, and they may submit requests 
for new VMs with low probability, the request process can be modelled as a Poisson 
process~\cite{trivedi:book:2001, mainkar:1996}. The same story holds for external request processes at both MSP and backend cloud.

The steps incurred in servicing a request in \textit{Macroservice Infrastructure} (MSI) are shown in 
the lower part of Fig. \ref{fig:si-rt}. User requests are submitted to a global finite queue, and then 
the scheduler processes them on an FCFS-basis. A request that finds the queue full will be rejected immediately. Once the 
request is admitted to the queue, it must wait until the scheduler in the Physical Machine 
Assigning Module (PMAM) processes it (first delay). 
As shown in Fig. \ref{fig:si-rt}, these steps are incorporated into the proposed performance model within the
Physical Machine Sub-Model (PMSM).
Once the request is assigned to one of the PMs, 
it will have to wait in that PM's input queue (second delay) until the VM Provisioning Module
(VMPM) instantiates and deploys the necessary VM (third delay) before the start of the actual service.
When a running request finishes, the capacity used by the corresponding VM is released and becomes 
available for servicing the next request.
These processes and delays are addressed in the Virtual Machine Sub-Model
(VMSM).

\begin{table}[!t]
  \renewcommand{\arraystretch}{1.1}
  \caption{Symbols and corresponding descriptions for microservice platform module.}
	\centering
       	\rowcolors{2}{white}{gray!25}
	\begin{tabular} {|c|p{6cm}|}
		\hline
		\textbf{Symbol} & \textbf{Description}	\\
		\hline
		MPM & Microservice Platform Module \\
		CSM & Container Sub-Model \\
		$\lambda$ & Mean arrival rate  to CSM\\
		$1/\alpha$ & Mean time that takes to obtain a VM \\
		$1/\beta$ &  Mean time that takes to release a VM \\
		$s$ & Min number of active VM in user's Host Group \\
		$S$ & Max size of the user's host group (VMs) \\
		$M$ & Max number of containers running on a VM\\ 
		$1/\mu$ & Mean value of container lifetime \\ 
		$\phi$ & The rate by which a container can be instantiated \\
		$u$ & Utilization in the user's Host Group \\ 
		$L_q$ & Microservice global queue size, $L_q$ = $S * M$  \\
		$bp_q$ & Blocking probability in the microservice global queue \\
		$\lambda_c$ & Arrival rate of requests for VMs originated from CSM \\
		$\eta_c$ & VM release rate imposed by CSM \\ 
		$\overline{wt_q}$ & Mean waiting time in the microservice global queue \\
		\hline
		\end{tabular}
	\label{tab:microsymbols}
\end{table}
		
\begin{table}[!t]
  \renewcommand{\arraystretch}{1.1}
  \caption{Symbols and corresponding descriptions for macroservice modules.}
	\centering
       	\rowcolors{2}{white}{gray!25}
	\begin{tabular} {|c|p{6cm}|}
		\hline
		\textbf{Symbol} & \textbf{Description}	\\
		\hline
		PMAM & Physical Machine Assigning Module \\
		PMSM & Physical Machine Sub-Model \\
		$\lambda_{x}$ & Mean value of external request to PMSM \\
		$\lambda_a$ & Aggregate arrival rate to PMSM that is equal to $\lambda_c + \lambda_x$ \\
		$B\!P_q$ & Blocking probability due to lack of room in the macroservice global queue \\
		$B\!P_r$ & Blocking probability due to lack of capacity in the cloud center \\
		$L_Q$ & Size of global input queues \\
		$1/\alpha$ & Mean lookup time in the pool \\
		$P_{\mbox{reject}}$ & Total probability of blocking ($B\!P_q+B\!P_r$) \\
		$P_s $ & Probability of successful search in pool \\
		$\overline{wt_Q}$ & Mean waiting time in macroservice global queue \\
		$\overline{lut}$ & Mean lookup delay among pools \\
		$\overline{{P\!M}_{wt}}$ & Mean waiting time in a PM queue \\
		\hline
		\hline
		VMPM & Virtual Machine Provisioning Module \\
		VMSM& Virtual Machine Sub-Model\\
		$m$ & Max number of VMs that a PM is set to host; $m$ is also the queue size in each PM \\ 
		$1/\eta$ & Mean value of VM lifetime \\ 
		$\delta$ & The rate by which a VM can be instantiated \\
		$\lambda_h$ & Arrival rate to a PM in the pool \\
		$N$ & Number of PMs in the servers pool \\
		$\overline{pt}$ & Mean VM provisioning time \\
		\hline
		$\overline{td}$ & Mean total delay imposed on requests in macroservice infrastructure \\
		\hline
		\end{tabular}
	\label{tab:macrosymbols}
\end{table}
To model the behavior of this system, we design three stochastic sub-models:
1) Container Sub-Model (CSM); 2) Virtual Machine Sub-Model (VMSM); and 3)
Physical Machine Sub-Model (PMSM).
These sub-models are also shown in Fig. \ref{fig:si-rt}.
Among the proposed sub-models, CSM captures the details 
of the microservice platform called Microservice Platform Module (MPM) and the other two (PMSM and VMSM) capture the 
details of the macroservice infrastructure, namely, Physical Machine Assigning Module (PMAM) and 
Virtual Machine Provisioning Module (VMPM).  We implement each module with a stochastic sub-model.
We combine all three stochastic sub-models and build an overall interactive performance model. 
Then, we solve this model to compute the cloud performance metrics: request rejection probability,
probability of immediate service and 
mean response delay as functions of variations in workload (request arrival rate), 
container lifetime, users quota (i.e., host group size), number of container per VM and system 
capacity (i.e., number of PMs in cloud center).
We describe our analysis in detail in the following 
sections, using the symbols and acronyms listed in Tables \ref{tab:microsymbols} and \ref{tab:macrosymbols}.

\section{Layered Analytical Model} \label{lam}
In this paper, we implement the sub-models using interactive Continuous Time Markov 
Chain (CTMC). These sub-models are interactive, such that the output of one model is input 
to the other ones and vice versa. Table \ref{tab:module-model} shows the modules and 
their corresponding stochastic sub-models, which will be discussed in detail in the following 
sections.
\begin{table}[!t]
  \caption{Modules and their corresponding sub-models.}
   \renewcommand{\arraystretch}{1.1}
	\centering
		\begin{tabular} {|l|c|c|}
		\hline
		\textbf{Component} & \textbf{Module} & \textbf{Stochastic sub-model} \\
		\hline
		Microservice Platform & MPM & CSM \\
		\hline
		Macroservice & PMAM & PMSM \\ \cline{2-3}
		Infrastructure  & VMPM & VMSM \\
		\hline
		\end{tabular}
	\label{tab:module-model}
\end{table}

\begin{figure}[!th]
\begin{center}
\includegraphics[width = 0.5\textwidth]{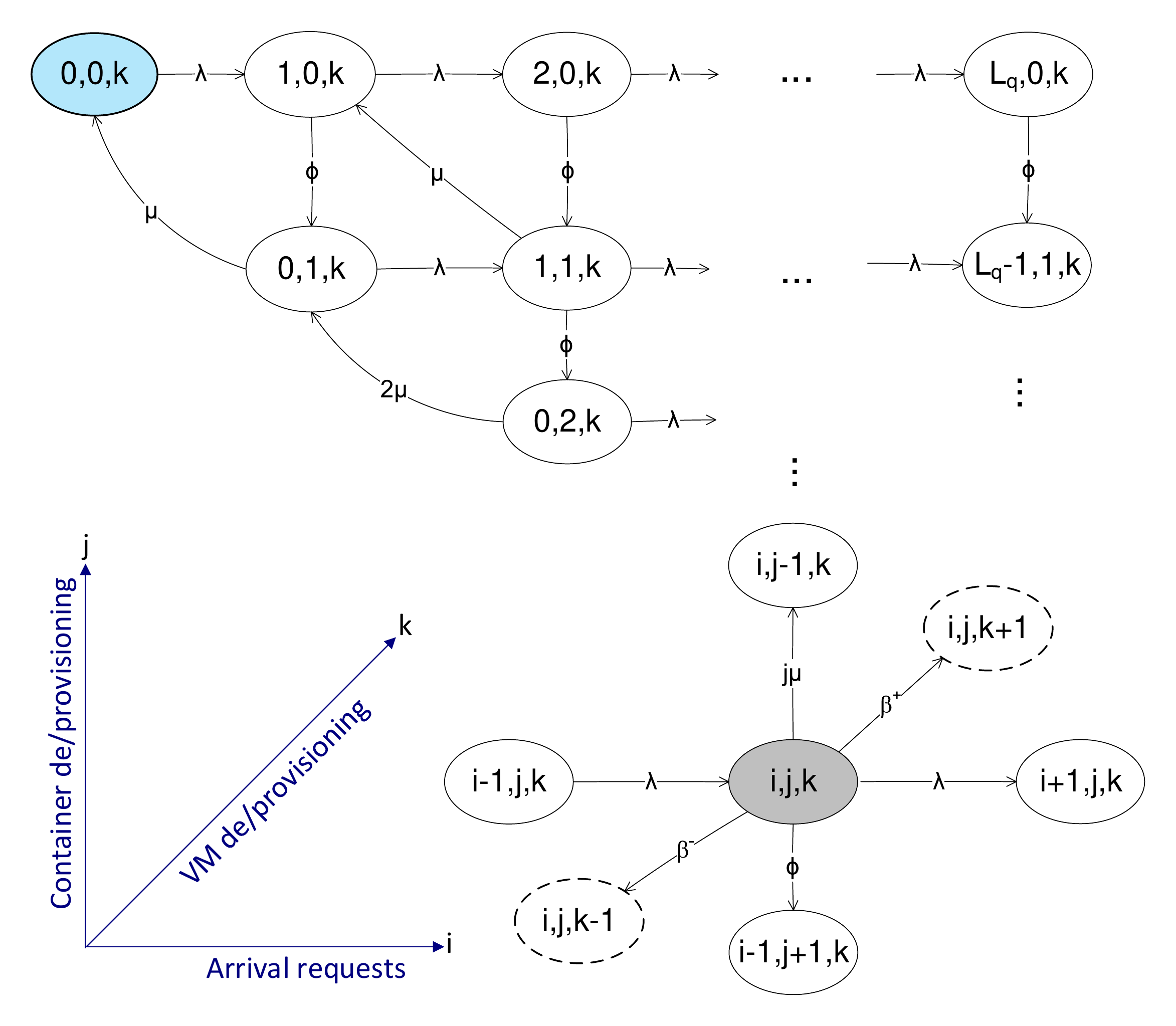}
\end{center}
\caption{Container Sub-Model (CSM).}
\label{fig:csm}
\end{figure}
\subsection{Microservice Platform Module} \label{mpm}
The container allocation process in the microservice platform is described in the Container Sub-Model
(CSM) shown in Fig. \ref{fig:csm}. CSM is a 3-dimensional CTMC with states that labelled as $(i,j,k)$ 
where $i$ indicates the number of requests in Microservice Global Queue, $j$ denotes the 
number of running containers in the platform, and $k$ shows the number of active VMs 
in the user's Host Group. In this work, we assume all inter-event periods are exponentially distributed. 
Each VM can accommodate up to $M$ containers which is set by the user. 
Since the number of potential users is high and a single user typically submits requests at a 
moment in time with low probability, the requests' arrival can be adequately modelled as a Poisson process 
\cite{grimmett} with rate $\lambda$. Let $\phi$ be the rate at which a container can be deployed 
on a VM and $1/\mu$ be the mean value of container lifetime (both exponentially distributed). 
So, the total service rate for each VM 
is the product of the number of running containers by $\mu$. Assume $\beta^+$ and $\beta^-$ are the 
rates at which the MSP can obtain and release a VM, respectively.

CSM asks for a new VM from 
backend cloud (i.e., macroservice infrastructure) when explicitly ordered by the MSP user or when the 
utilization of the host group is equal or greater than a predefined value. For state $(i,j,k)$, utilization 
($u$) is defined as follows,
\begin{equation}
u = \frac{i+j}{k \times M}
\end{equation}
in which $M$ is the maximum number of containers that can be run on a single VM.

The value of 
$u$ indicates the ratio of active containers to the total available containers at each state.
In this scenario, if the utilization drops lower than a predefined threshold, the CSM will release one VM to 
optimize the cost. A VM can be released if there are no running containers on it, so the 
VM should be fully \textit{decommissioned} in advance. Also, the CSM holds a minimum number 
of VMs in the Host Group regardless of utilization, in order to maintain the availability 
of the service ($s$). The user may also set another value for its application(s)
($S$) indicating that the MSP can not request more than $S$ VMs from macroservice infrastructure 
on behalf of the user. Thus, the application scales up at most to $S$ VMs in case of high traffic 
and scales down to $s$ VMs in times of low utilization. We set the global queue size ($L_q$) to 
the total number of containers that it can accommodate at its full capacity 
(i.e., $L_q$ = $S \times M$).  Note that the request will be blocked if
the user reached its capacity, regardless of the Global Queue state. 

State $(0,0,s)$ indicates that there are no requests in the queue, no running container, and the 
Host Group consists of $s$ VMs which is the minimum number of VMs that the user maintains in its Host Group. 
Consider an arbitrary state such as $(i,j,k)$, in which five transitions might happen:
\begin{enumerate}
\item Upon the arrival of a new request, the system moves to state $(i+1,j,k)$ with a rate of $\lambda$  
if the user still has the capacity (i.e., $i+j < S \times M$), otherwise the request will be 
blocked, and the system stays in the current state. 
\item The CSM instantiates a container with rate $\phi$ for the request in the head of Global Queue and moves to $(i-1,j+1,k)$.
\item The lifetime of a container is exponentially distributed and ends with a rate of $j\mu$ and the system moves to $(i,j-1,k)$. 
\item If the utilization goes higher than the threshold, MSP requests a new VM, and the system moves
to state $(i,j,k+1)$ with a rate of $\beta^+$. 
\item Or the utilization drops below a certain threshold, MSP decommissions a VM, and the system releases the idle VM, so it 
moves to $(i,j,k-1)$ with rate $\beta^-$.
\end{enumerate}

Note that CSM (depicted in Fig. \ref{fig:csm}) is only for one user (or one application in another sense). 
In this work, we assume homogeneous users so that we only need to solve one CSM regardless of the number of users at MSP. 
Suppose that $\pi_{(i,j,k)}$ is the steady-state probability for the CSM (Fig. \ref{fig:csm}) to be in the state $(i,j,k)$, which is calculated the same way as in~\cite{hamzeh-tpds}. So, the blocking probability in 
CSM can be calculated as follows:
\begin{equation} \label{eq:bp-csm}
bp_q = \sum \pi_{(i,j,k)}  \quad \text {if i + j = $L_q$} 
\end{equation}
 We are also interested in two probabilities by which the MSP requests ($p_{\text{req}}$) or 
 releases ($p_{\text{rel}}$) a VM.
\begin{align} 
p_{\text{req}} = & \sum \pi_{(i,j,k)}  \quad \text {if} \quad  u \geq \text{high-util}  
\;\; \& \;\; k < S \\
p_{\text{rel}} = & \sum \pi_{(i,j,k)}  \quad \text {if} \quad u \leq \text{low-util} 
\;\; \& \;\; k > s  
\end{align} 
Using these probabilities, the rate by which microservice platform requests 
($\lambda_{c}$) or releases ($\eta_c$) a 
VM can be calculated.
\begin{align} \label{eq:lambda-theta}
\lambda_{c} = & \: \lambda \times p_{\text{req}}   \\
\eta_c = & \: \mu \times p_{\text{rel}}  
\end{align}
In order to calculate the mean waiting time in queue, we first calculate the number of requests in 
the queue as
\begin{equation} \label{eq:q-ave-mic}
\overline{q} = \sum_{i=0}^{L_q} i \cdot \pi_{(i,j,k)}
\end{equation}
Applying Little's law \cite{kleinrock}, the mean waiting time in the global queue
is given by:
\begin{equation} \label{eq:wt-ave-mic}
\overline{wt_q} = \frac{\overline{q}}{\lambda(1-bp_q)}
\end{equation}
Instantiation time for containers ($1/\phi$) will be added to $\overline{wt_q}$ to 
obtain the total delay in the microservice platform.

CSM has interactions with both the physical machine provisioning sub-model
(i.e., PMSM, described in section \ref{pmsm}) and the virtual machine provisioning sub-model 
(i.e., VMSM, described in section \ref{vmsm}). $\lambda_{c}$ and $\eta_c$ are used by 
PMSM and VMSM, respectively. The details of interactions among sub-models will 
be explained fully in section \ref{inter}.

\subsection{Macroservice Infrastructure Model}
\subsubsection{PM Provisioning Sub-Model} \label{pmsm} 
\begin{figure}[!th]
\begin{center}
\includegraphics[width = 0.45\textwidth]{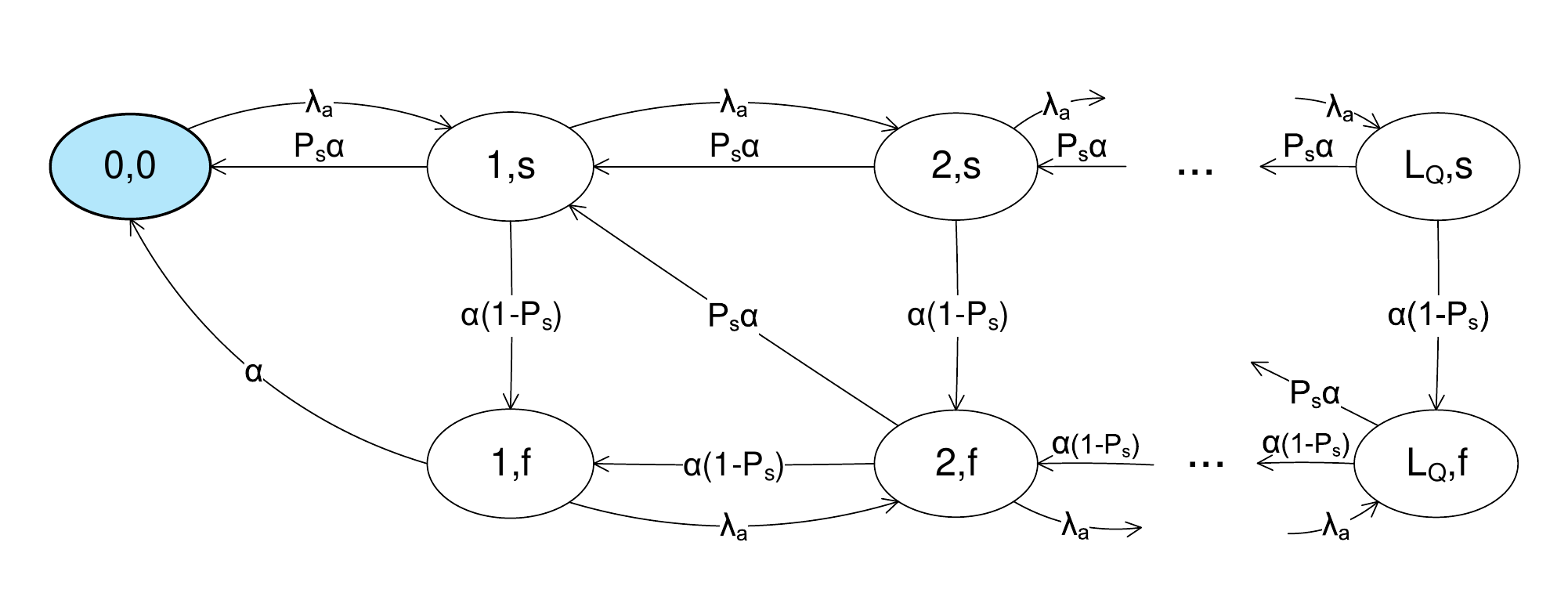}
\end{center}
\caption{Physical Machine Sub-Model (PMSM).}
\label{fig:pmsm}
\end{figure}

The resource allocation process is described in the Physical Machine Sub-Model (PMSM) 
shown in Fig. \ref{fig:pmsm}. PMSM is a 2-dimensional CTMC (i.e., inter-event epochs are exponentially distributed) 
that records the number of requests in the global queue and the latest state of provisioning. The state $(i,s)$ indicates that the
last provisioning was successful while $(i,f)$ shows that the last provisioning has been failed.
We assume that the mean arrival rate is $\lambda_a$, and the global queue size is $L_Q$. 
One more request can be at the deployment unit for provisioning; thus, the capacity of 
the system is $L_{Q}+1$.

Let $P_s$ be the success probability of finding a PM that can accept the current request 
in the pool. We assume that $1/\alpha$ is the mean lookup delay for finding an appropriate
PM in the pool that can accommodate the request. Upon the arrival of the first request, the system moves 
to state $(1,s)$. Afterwards, depending on the upcoming event, three possible transitions can occur: 
\begin{enumerate}[(a)]
\item Upon the arrival of a new request, the system transits to state $(2,s)$ with rate $\lambda_a$. 
Note that this request might arrive from two sources: 1) directly from macroservice users; or 
2) from microservice platform due to overutilization.
See the inputs to the global queue in the lower part of Fig. \ref{fig:si-rt}.
\item Or, a PM in the pool accepts the request so that system moves back to the state $(0,0)$ with 
rate $P_s\alpha$. 
\item Or none of the PMs in the pool can accept the request so the system 
moves to state $(1,f)$ with rate $(1-P_s)\alpha$. This way, the scheduler gives another chance to 
the request for provisioning; if the second attempt doesn't go through, the request will be rejected due
to lack of capacity. From states $(i, f)$, for $i>0$ two moves are possible: 1) the system goes to $(i-1,s)$ if
the provisioning was successful or 2) it moves to $(i-1,f)$ if the provisioning has failed for the second time. 
At state $(1,f)$ the system moves back to $(0,0)$ regardless of the previous provisioning state. 
Note that the number of retry attempts here is set to 2 and can be adjusted in the cloud service 
controller~\cite{savi}. 
\end{enumerate}

In this sub-model the lookup rate ($\alpha$) and macroservice users' request rate are exogenous 
parameters, microservice users' requests rate ($\lambda_c$) is calculated from CSM and success 
probability ($P_s$) is calculated from the VMSM which will be discussed in next section. 

\begin{figure}[!ht]
\begin{center}
\includegraphics[width=0.6\columnwidth]{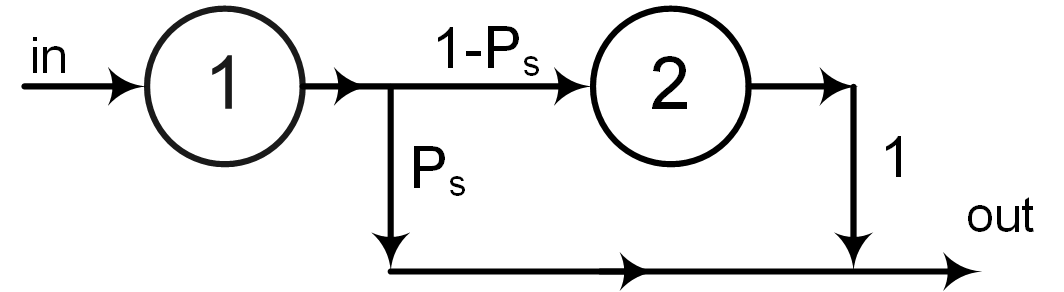}
\end{center}
\caption{Two steps of look-up delay.}
\label{fig:coxian-st}
\end{figure}

Using steady-state probabilities $\pi_{(i)}$, blocking probability can be calculated. 
Requests may experience two kinds of blocking events:\\
\begin{enumerate} [(a)]
\item Blocking due to a full global queue occurs with the probability of
\begin{equation} \label{eq:BP_q}
B\!P_q = \pi_{(L_Q,s)} + \pi_{(L_Q,f)}  
\end{equation}
\item Blocking due to insufficient resources (PMs) at server pools occurs with the probability of
\cite{heymen-sobel} 
\begin{equation} \label{eq:BP_r}
B\!P_r = \sum_{i=1}^{L_Q} \frac{\alpha(1-P_s)}{\alpha P_s + \lambda_a} \pi_{(i,f)}  
\end{equation}
\end{enumerate}
\noindent Eq. \ref{eq:BP_r} is the ratio of aggregate rates by which the system blocks requests
due to insufficient resources to all other rates over all states.
The probability of reject is, then, $P_{reject} = B\!P_q + B\!P_r$.
To calculate the mean waiting time in queue, we first calculate the mean
number of requests in the queue as
\begin{equation} \label{eq:q-ave}
    \overline{q} = \sum_{i=1}^{L_Q} i (\pi_{(i,s)} + \pi_{(i,f)})
\end{equation}
Applying Little's law \cite{kleinrock}, the mean waiting time in the global queue
is given by (first delay):
\begin{equation} \label{eq:wt-ave}
\overline{wt_Q} = \frac{\overline{q}}{\lambda_{a}(1-B\!P_q)}
\end{equation}
Lookup time for appropriate PM can be considered as a Coxian distribution with two steps 
(Fig. \ref{fig:coxian-st}). 

\noindent Therefore according to \cite{trivedi:book:2001}, the look-up time (second delay)
can be calculated as follows.
\begin{equation} \label{eq:lut-ave}
\overline{lut} = \frac{1/\alpha+ (1-P_s)(1/\alpha)}{1-B\!P_q}
\end{equation}

\subsubsection{VM Provisioning Sub-Model} \label{vmsm}
\begin{figure}[!t]
\begin{center}
\includegraphics[width = 0.49\textwidth]{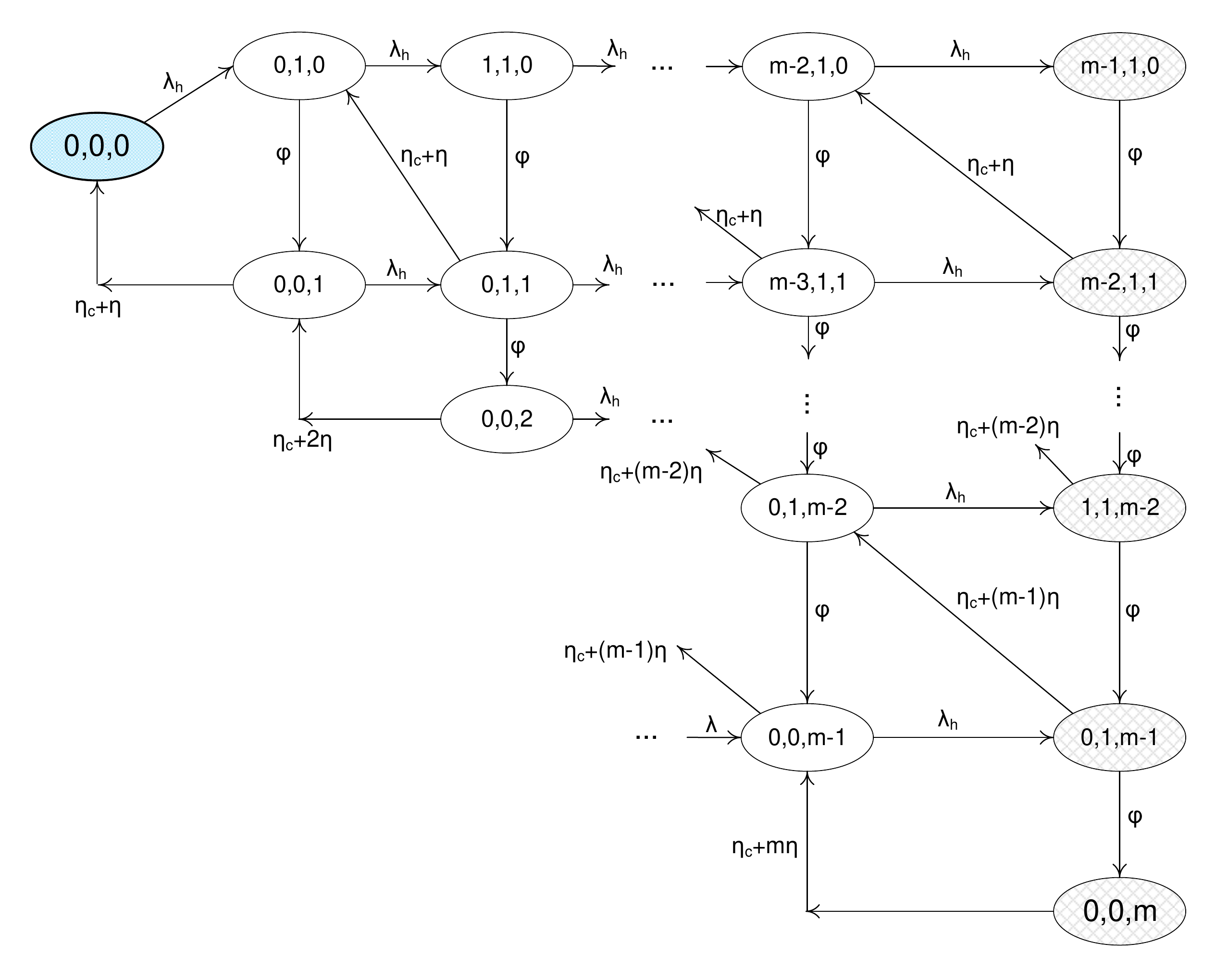}
\end{center}
\caption{Virtual Machine Provisioning Sub-Model for a PM in the pool (VMSM).}
\label{fig:vmsm}
\end{figure}
Virtual Machine provisioning Sub-Model (VMSM) captures the instantiation, deployment and
provisioning of VMs on a PM. VMSM also incorporates the actual servicing of each request (VM)
on a PM. Fig.~\ref{fig:vmsm} shows the VMSM (a CTMC) for a \emph{single} PM in the servers' 
pool. As we assume homogeneous PMs, all VMSMs for the PMs are identical in terms of arrival 
and instantiation rates. Consequently, the server pool is modelled with a set of identical 
VMSMs so that we only need to solve one of them. 

   
Each state in Fig. \ref{fig:vmsm} is labelled by $(i,j,k)$ in which $i$ indicates the number of 
requests in PM's queue, $j$ denotes the number of requests that are under provisioning by the 
hypervisor and $k$ is the number of VMs that are already deployed on the PM. Note that we set 
the queue size at each PM to $m$, the maximum number of VMs that can be deployed on a PM. 
Also, the hypervisor can only deploy one VM at a time to the PM, so the value of $j$ is 0 when the
instantiation unit is idle and will be 1 when it is deploying a VM. Let $\delta$ be the 
rate at which a VM can be deployed on a PM and $\eta$ be the service rate of each 
VM. So, the total service rate for each PM is the product of the number of running VMs by $\eta$.
Note that in VMSM, a VM may get released due to two events; first, the service time of 
VM is finished (the rate is $\eta$) and second if the microservice platform specifically asks 
for termination of that VM (with a rate of $\eta_c$, see Fig. \ref{fig:vmsm}).
 
State $(0,0,0)$ indicates that the PM is empty, and there is no request either in the queue or 
in the instantiation unit. Upon the arrival of a request, the model transits to state $(0,1,0)$. 
The arrival rate to each PM is given by:
\begin{equation}\label{eq:arr-rate-hot}
\lambda_h = \frac{\lambda_a(1-B\!P_q)}{N}
\end{equation}
in which $N$ is the number of PMs in the pool. Note that $B\!P_q$, used in (\ref{eq:BP_q}), is 
obtained from PMSM. The state transition in VMSM can occur due to request arrival, VM 
instantiation or service completion. From state $(0,1,0)$, the system can move to state 
$(1,1,0)$ with rate $\lambda_h$. From $(1,1,0)$, the system can transit to $(0,1,1)$ with 
rate $\delta$ (i.e., instantiation rate) or upon arriving a request moves to state $(2,1,0)$. 
From $(0,1,1)$, system can move to $(0,0,2)$ with rate $\delta$, transits to $(0,1,0)$ with rate 
$\eta + \eta_c$ (i.e., service completion or an explicit request from microservice platform),
or again upon arriving a request, system can move to state $(1,1,1)$ with rate $\lambda_h$. 

Suppose that $\pi_{(i,j,k)}$ is the steady-state probability for the PM model (Fig. \ref{fig:vmsm}) 
to be in the state $(i,j,k)$. Using steady-state probabilities, we can obtain the 
probability that at least one PM in the pool can accept the request for provisioning. 
At first, we need to compute the probability that a 
PM cannot admit a request for provisioning ($P_{na}$): 
\begin{equation} 
P_{na} = \sum \pi_{(i,j,k)} \quad \text{if} \; i+j+k = m
\end{equation} 
Therefore, the probability of successful provisioning ($P_s$) in the pool can be obtained as
\begin{equation} \label{eq:pro-succ-hot}
P_s = 1 - (P_{na})^{N}
\end{equation} 
Note that $P_s$ is used as an input parameter in the PMSM (Fig. \ref{fig:pmsm}). 

From VMSM, we can also obtain the mean waiting time at PM's queue (third delay: 
$\overline{{P\!M}_{wt}}$) and mean provisioning time (fourth delay: $\overline{pt}$) by using 
the same approach as the one that led to Eq. (\ref{eq:wt-ave}). As a result, the total delay 
in macroservice infrastructure before having VM ready for servic is given by:
\begin{equation} \label{eq:total-delay}
\overline{td} = \overline{wt_Q} + \overline{lut} + \overline{{P\!M}_{wt}}+ \overline{pt} 
\end{equation} 
Using Eq. \ref{eq:total-delay} we can calculate $\alpha$ and $\beta$ rates that are input 
parameters for the container sub-model (CSM). We assume that, without loss of generality, 
the amount of time that takes to obtain a VM is equal to the time that is needed to release the VM.  
\begin{equation} \label{eq:alpha-beta}
\alpha = \beta = 1/ \overline{td}
\end{equation}

\subsection{Interaction among Sub-Models}\label{inter}
\begin{figure}[!t]
\begin{center}
\includegraphics[width = 0.65\columnwidth]{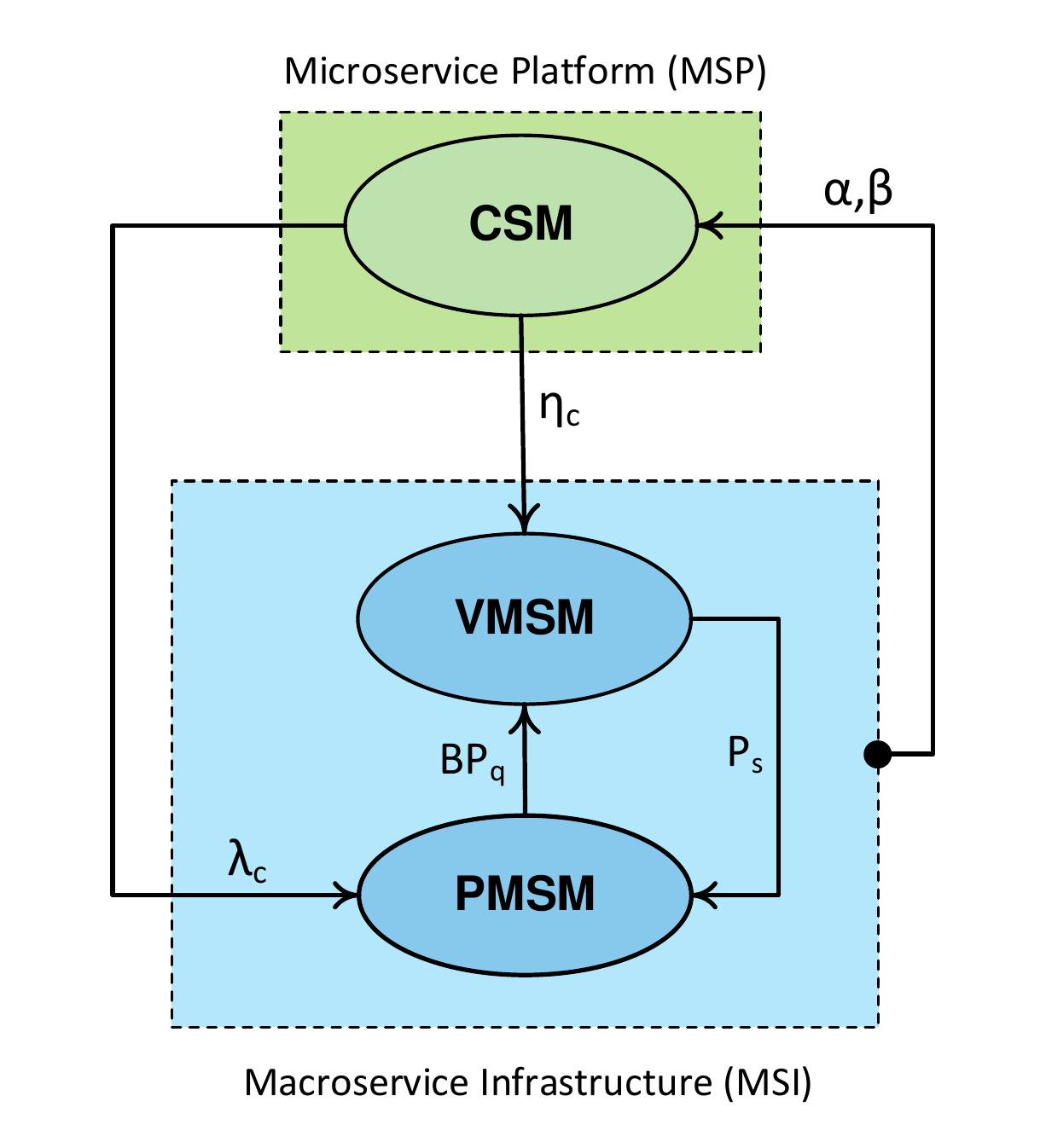}
\end{center}
\caption{Interaction diagram among layers; microservice platforms have been modelled
with one sub-model while macroservice infrastructure has been modelled with two sub-models.}
\label{fig:inter}
\end{figure}

The interactions among sub-models are depicted in Fig. \ref{fig:inter} and also described 
in Table \ref{tab:sub-model-output}. From container sub-model (CSM), the request rate for 
new VMs (i.e., $\lambda_{c}$), as well as release rate of VMs (i.e., $\eta_c$), can be calculated 
(see Eq. \ref{eq:lambda-theta}); these are used as inputs to PM assigning sub-model 
(PMSM) and VM provisioning sub-model (VMSM), respectively. From the macroservice model, 
which includes both PMSM and VMSMs, the amount of time that CSM can obtain 
(i.e., $1/\alpha$) or release (i.e., $1/\beta$) a VM will be calculated; these two are input 
parameters for CSM. The VMSMs compute the steady-state success probability ($P_s$ that at 
least a PM in the pool can accept the request. This success probability is used as input 
parameters to the PMSM. The PMSM computes the blocking 
probability, $B\!P_q$, which is the input parameter to the VM provisioning sub-model.
Since $\alpha$ and $\beta$ are computed using both PMSM and VMSM, we show them as the 
outputs of the MSI in Fig. \ref{fig:inter} as well as Table \ref{tab:sub-model-output}.
\begin{table}[!ht]
  \caption{Sub-models and their inputs/outputs.}
   \renewcommand{\arraystretch}{1.1}
	\centering
       	\rowcolors{2}{white}{gray!25}
		\begin{tabular} {|c|c|c|}
		\hline
		\textbf{Model or Sub-model} & \textbf{Input(s)} & \textbf{Output(s)} \\
		\hline
		CSM & $\alpha, \beta$ & $\lambda_{c}$, $\eta_c$\\
        \hline
	        PMSM & $\lambda_{c},P_s$ & $B\!P_q$\\
		VMSM & $\eta_c,B\!P_q$ & $P_s$ \\
		\hline
		\end{tabular}
	\label{tab:sub-model-output}
\end{table} 

As can be seen, there is an inter-dependency among sub-models. This cyclic dependency is resolved 
via the fixed-point iterative method \cite{mainkar:1996} using a modified version of the successive 
substitution approach.

For numerical experiments, the successive substitution method (Algorithm \ref{alg:ssm}) is 
continued until the difference between the previous and current value of blocking probability 
in the global queues for both CSM and PMSM (i.e., $B\!P_q$ and $bp_q$) are less than $10^{-6}$ 
as the \emph{max\_err}. 
The integrated model usually converges to the solution in less than 15 
iterations for each while-loop. When Algorithm \ref{alg:ssm} converges, then performance metrics 
will be calculated. A \textit{null} return value has been used to signal the algorithm has not converged and thus needs a better initial value.

It can be shown that fixed-point variable $BP_q$ can be expressed as a function of $P_s$, and variable
$bp_{q0}$ can be expressed as a function of $\alpha$ and $\beta$. 
For this reason, before starting the fixed-point iteration, we guess an initial value for $P_s$, and we check the values of $BP_q$ in successive iterations to determine the condition for convergence in the inner loop.
The same procedure will be followed for the external $while$ loop, and finally, the fixed-point iteration will converge. Fixed-point iteration starts by calling the function CSM(·) to obtain the steady solution of the container sub-model. The output of CSM(·) is the steady-state queue size ($bp_{q0}$) that the job will be blocked in the microservice platform. Function PMSM(·) uses this updated $bp_{q0}$ as an input parameter and solves the PMSM sub-model for the steady-state solution. Then, the inner loop will iterate until we get a steady-state
solution for the macroservice model. Afterwards, $\alpha$ and $\beta$ will be calculated using Eq. 20. The steady-state 
value for queue size at the microservice platform $bp_{q1}$ is obtained and compared to the previous value.
If the difference is less than the threshold, then the algorithm converges, and the final steady-state values 
for $BP_{q0}$ and $bp_{q0}$ will be calculated. The proof of the existence of a solution, for this method in general,
has been detailed in ~\cite{ghosh:2012}.

\begin{algorithm}[!htb]
\Input {Initial success probability in the pool: $P_{s}$}
\Input {Initial rate for requesting or releasing a VM: $\alpha, \beta$}
\Output {Blocking probability in Global Queues: $B\!P_q, bp_q$}
counter\_a $\leftarrow$ 0; max\_a $\leftarrow$ 10; diff\_a $\leftarrow$ 1\; 
counter\_b $\leftarrow$ 0; max\_b $\leftarrow$ 10; diff\_b $\leftarrow$ 1\;
$bp_{q0}$ $\leftarrow$ CSM ($\alpha, \beta$); ({\color{blue} Eq. \ref{eq:bp-csm}})\\

	\While{diff\_a $\geq$ max\_err}
	{
		counter\_a $\leftarrow$ counter\_a + 1\;
		$B\!P_{q0}$ $\leftarrow$ PMSM ($bp_{q0}, P_{s}$); ({\color{blue} Eq. \ref{eq:BP_q}})\\
		\While{diff\_b $\geq$ max\_err}
		{
			counter\_b $\leftarrow$ counter\_b + 1\;
			$P_{s}$ $\leftarrow$ VMSM ($B\!P_{q0}$)\; 
			$B\!P_{q1}$ $\leftarrow$ PMSM ($bp_{q0}$,$P_{s}$)\;
			$\mbox{diff\_b} \; \leftarrow \; \mid\!B\!P_{q1}-B\!P_{q0}\!\mid$\;
			$B\!P_{q0} \leftarrow B\!P_{q1}$\;
			\tcp{if maximum try is reached}
			\If{counter\_b = max\_b}
			{
				\Return null;
			}
		}
		$[\alpha, \beta]$ $\leftarrow$ calculate\_alpha\_beta(); ({\color{blue} Eq. 
        \ref{eq:alpha-beta}})\\
		$bp_{q1}$ $\leftarrow$ CSM ($\alpha, \beta$)\;
		$\mbox{diff\_a} \; \leftarrow \; \mid\!bp_{q1}-bp_{q0}\!\mid$\;
		$bp_{q0} \leftarrow bp_{q1}$\;
		\If{counter\_a = max\_a}
		{
			\Return null;
		}
	}
\Return $[B\!P_{q0}, bp_{q0}]$;	
\caption{Successive Substitution Method.}
\label{alg:ssm}
\end{algorithm}

\subsection{Scalability and flexibility of the integrated model} \label{scale}
Both macroservice infrastructure and microservice platform may scale up or down in 
terms of global queue sizes, number of PMs in the pool, number of VMs in user quota and the 
degree of virtualization (in both VM and container level) which are referred as 
\textit{design parameters}. Table \ref{tab:scale} shows the relationship between 
the number of states in each sub-model and their associated design parameters. 
\begin{table}[!ht]
 \renewcommand{\arraystretch}{1.1}
\caption{Relationship between the size of sub-models and design parameters.}
\centering
 \begin{tabular}{|c|c|p{3.5cm}|}
	\hline
	\textbf{Sub-Model} & \textbf{Design Parameters} & \textbf{No. of States: f(i)} \\
	\hline
	CSM &  $s,S,M$ & f($s$,$S$,$M$) =  $MS^2$ - $sMS$ \\
	\hline
	PMSM & $L_Q$ & f($L_Q$) = 2$L_Q$ + 1 \\
	\hline
	\multirow{3}{*}{VMSM} &  & f(m) = 3, \quad \quad \quad if m=1 \\ 
	 & m & f(m) = 6, \quad \quad \quad if m=2 \\ 
	 &  & f(m) = $2f(m-1)-f(m-2)+1$, \quad \quad \quad if m$>$2 \\ 
	\hline
 \end{tabular}
\label{tab:scale}
\end{table}

As can be seen from Table \ref{tab:scale}, the number of states in each sub-model has a 
linear or polynomial dependency to design parameters that guarantees the scalability of 
the integrated model. Note that VMSMs are identical for all PMs in the pool. Therefore, 
we only need to solve one VMSM regardless of the number of PMs in the macroservice 
infrastructure.

\section{Numerical Validation} \label{env}
In this section, we first describe our experimental setup on the Amazon EC2 cloud; then, we discuss the results and insights 
that we obtained from the real implementation and deployment of our microservice platform. Next, the analytical model 
will be tuned and validated against experimental results. Finally, we leverage the analytical model to study and investigate
the performance of the microservice platform at a large scale for various configuration and parameter settings.   

\subsection{Experimental Setup} \label{esr}
Here, we present our microservice platform and discuss experiments that we have 
performed on this platform. For experiments, we couldn't use available third-party platforms such as 
Docker Cloud or Nirmata as we needed full control of the platform for monitoring, parameter setting, 
and performance measurement. As a result, we have created a microservice platform from scratch 
based on the conceptual architecture presented in Fig. \ref{fig:micro-concept-model}.
We employed Docker Swarm as the cluster management system, Docker as the container engine 
and Amazon EC2 as the backend public cloud. We developed the front-end in Java for the microservice 
platform that interacts with the cluster management system (i.e., Swarm) through REST APIs. 
The microservice platform leverages three initial VMs with two configured in 
\emph{worker} mode and another one in \emph{master} mode to manage the Docker Swarm cluster. 
All VMs are of type \texttt{m3.medium} (1 virtual CPU with 3.5\,GB memory). In our deployment, 
we have used \emph{Consul} as the discovery service that has been installed on the Swarm Manager VM.

For the containers, we have used Ubuntu 16.04 image available on Docker Hub. Each running 
container was restricted to use only 512\,MB memory, thus making the capacity of a VM to 
be seven containers. The Swarm manager strategy for distributing containers on 
worker nodes was \emph{binpack}. The advantage of this strategy is that fewer VMs can 
be used since Swarm will attempt to put as many containers as possible on the current 
VMs before using another VM. Table \ref{tab:si-range-exp} presents the input values 
for the experiments.

\begin{table}[!ht]
\caption{Range of parameters for experiments.}
\renewcommand{\arraystretch}{1.1}
\centering
\rowcolors{1}{white}{gray!25}
\begin{tabular}{|p{3cm}|c|c|}
\hline
\textbf{Parameter} & \textbf{Value(s)} & \textbf{Unit} \\ 
 \hline
Arrival rate &  $20\upto 40$ & req/min \\
Queue size & 10 & container \\
VM capacity & 7 & container \\
Container lifetime & 2 & minute \\
Desired utilization & $70\%\upto90\%$ & N/A \\
Initial cluster size & 2 & VM \\
Max cluster size & 10 & VM \\
\hline
\end{tabular}
\label{tab:si-range-exp}
\end{table}

In order to control the experiment's costs, we have limited the cluster size to a maximum of
10 running VMs for the application, which gives us a maximum capacity of 70 running containers.
For the same reason, we set the container lifetime as 2 minutes in the experiment.
Under this configuration, our experiment takes up to 640 minutes. The results of our experiment are 
presented in Fig. \ref{fig:exp}. Note that the $X$ axis in Fig.~\ref{fig:exp} is experiment 
time in which we report the average values of performance indicators in every single minute;
hereafter, we call each minute of the experiments an \textit{iteration}. As can be seen in Fig~\ref{fig:exp},
the result of the experiment from iteration 60 to 580 has been ignored to present the interesting events 
and moments during the experiment more clearly on graphs\footnote{Another version of Figure \ref{fig:exp} including
the whole experiment and utilization plot can be found \href{http://pacs.eecs.yorku.ca/exfiles/Fig8-V2.pdf}{\color{blue}{\underline{here}.}}}.  

In the experiment, the lower and upper utilization thresholds are set to 
70\% and 90\%, respectively (shaded area in the fourth plot of
the full version of Fig.~\ref{fig:exp} linked in the footnote
that shows the areas where the cluster is underloaded or overloaded). The arrival rate has a Poisson distribution with a mean value of 20 to 40 requests per minute, shown in the second plot of 
Fig.~\ref{fig:exp} with a blue line.
In the first plot, the red line shows the number of running VMs and the blue line enumerates the number of running containers.

\subsection{Experimental Results} \label{er}
\begin{figure}[!t]
\begin{center}
\includegraphics[width=0.95\columnwidth]{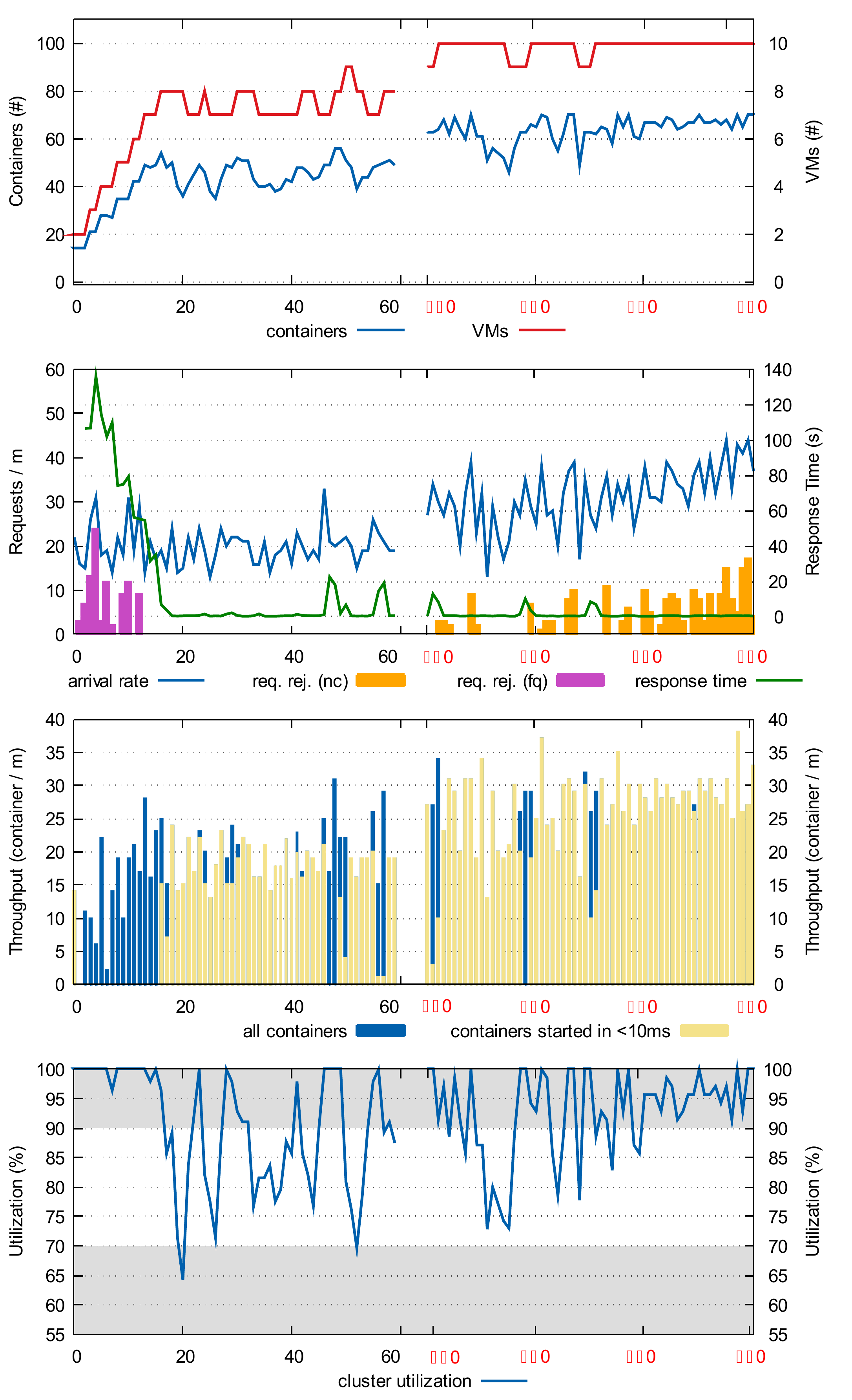}
\end{center}
\caption{Experimental results; see Table \ref{tab:si-range-exp} for parameter settings.}
\label{fig:exp}
\end{figure}

An interesting observation here is the behaviour of the system at the beginning of the experiment. 
The capacity of the cluster is 2 VMs that can support up to 14 containers. The experiment starts with 20 requests per minute, for 
which the current capacity is not enough. Therefore, as can be seen from iteration 0 to 20, the number of VMs and containers 
increases linearly (first plot in Fig.~\ref{fig:exp}). This is due to the high utilization of the cluster (fourth plot in the full version of
Fig.~\ref{fig:exp} linked in the footnote) that triggers the autonomic manager to scale up the cluster. At the same time, in the second plot in
Fig.~\ref{fig:exp}, the response time is declining after a sudden jump (i.e., up to approximately 140\,s) and also we see 
many rejected requests due to \emph{full queue} (indicated by legend ``req. rej. (fq)'') during the first 20 iteration 
of the experiment. 

In the third plot of Fig.~\ref{fig:exp}, we show the throughput of the system by measuring the number of successful 
containers that have been provisioned during each iteration in the experiment. Also, for each iteration, we show the 
percentage of containers that have been provisioned without queuing by yellow colour. In our experiment, if a request 
gets service in less than 10\,ms, we categorize it as \emph{immediate service}, which indicates that the request did not 
experience any queuing in the system. As can be seen, during the first 20 iterations of the experiment, almost all 
of the requests for containers have experienced a delay due to queuing. 

Around iteration 20, the system reaches the capacity that can handle the workload so we can see that response time 
drops to less than a second (approximately 450\,ms), there is no blocking request due to full queue anymore, and 
utilization is back to the normal area. Also, we can see that after iteration 20 most of the request gets immediate 
service so that containers are being provisioned for customers without any delay (look at yellow bars in the 
third plot that are indicating this fact). We let the experiment continue from iteration 60 to 580 while 
increasing the workload smoothly. During this time, the system adapts to the changing workload accordingly and 
maintains the performance indicators at desired ranges.

However, after iteration 580, we can see that (first plot in Fig.~\ref{fig:exp}) the cluster is about its 
capacity (i.e., 10 VMs and 70 containers). As a result, we notice some request rejections due to 
\emph{no capacity} in the system (indicated with a legend of ``req. rej. (nc)'' in the second plot). 
When the workload gets beyond the 30 requests per second (i.e. after iteration 600), the cluster gets its 
full capacity, which leads to continued rejection of requests due to lack of capacity. Despite 
request rejection and running at full capacity, performance indicators (i.e., response time and 
immediate service) are desirable as opposed to what we have witnessed at the beginning of the 
experiment. In the last 20 iterations of the experiment since the system is running at capacity, the 
autonomic manager drops the new requests immediately so that the queue gets cleared very fast; 
therefore, when capacity becomes available, the new request will get into the service immediately 
which results in very good response time as well. However, at the beginning of the 
experiment, i.e. first 20 iterations, the autonomic manager knows that there is extra capacity, 
so it lets the requests to be queued till the new VMs get provisioned; VM provisioning takes 
around 110\,seconds on average which contributes to the long response times at the beginning of the 
experiment.  
 
\subsection{Validation of the Analytical Model}
In this section, we validate the analytical model with the results of the experiments presented
in section \ref{er}. We use the same parameters, outlined in Table \ref{tab:si-range-exp}, for both experiments 
and numerical analysis. The analytical model has been implemented and solved in Python using
NumPy, SciPy and Sympy libraries \cite{python}. Table \ref{tab:comp} shows the 
comparison between the results from the analytical model and experiment. 
As can be seen, both the analytical model and experimental results are well in tune with 
an error of less than 10\%. Note that in the experiment, we consider requests that get into
service less than 10\,ms as immediate service; This value has been approximated based on
our experience with SAVI~\cite{savi} cloud in processing requests when there is no queuing. It might 
be different in other cloud data centers. It should be noted that 10 ms is the request processing and not
the resource provisioning. Changing this value will directly impact the resulted number of requests with immediate
service in our calculation in both analytical and experimental evaluation.  
 
\begin{table}[!ht]
\caption{Corresponding results from the analytical model and experiment.}
\renewcommand{\arraystretch}{1.3}
\centering
\rowcolors{1}{white}{gray!25}
\begin{tabular}{|p{3cm}|c|c|c|}
\hline
\textbf{Parameter} & \textbf{Analytical Model} & \textbf{Experiment} \\ 
\hline
Response Time  & 2.89 & 3.211  \\
Utilization  & 81.4\% & 85.3\%\\
Mean No of VMs & 7.12 & 7.61 \\
Mean No of Containers & 39.8 & 43.8 \\
Immediate Service Prob. & 0.7415 & 0.782 \\
\hline
\end{tabular}
\label{tab:comp}
\end{table}

\begin{figure*}[!t]
\begin{center}
\subfigure[Request rejection probability in microservice platform.]
{\label{fig:rp-micro-first}
\includegraphics[width = 0.45\textwidth]{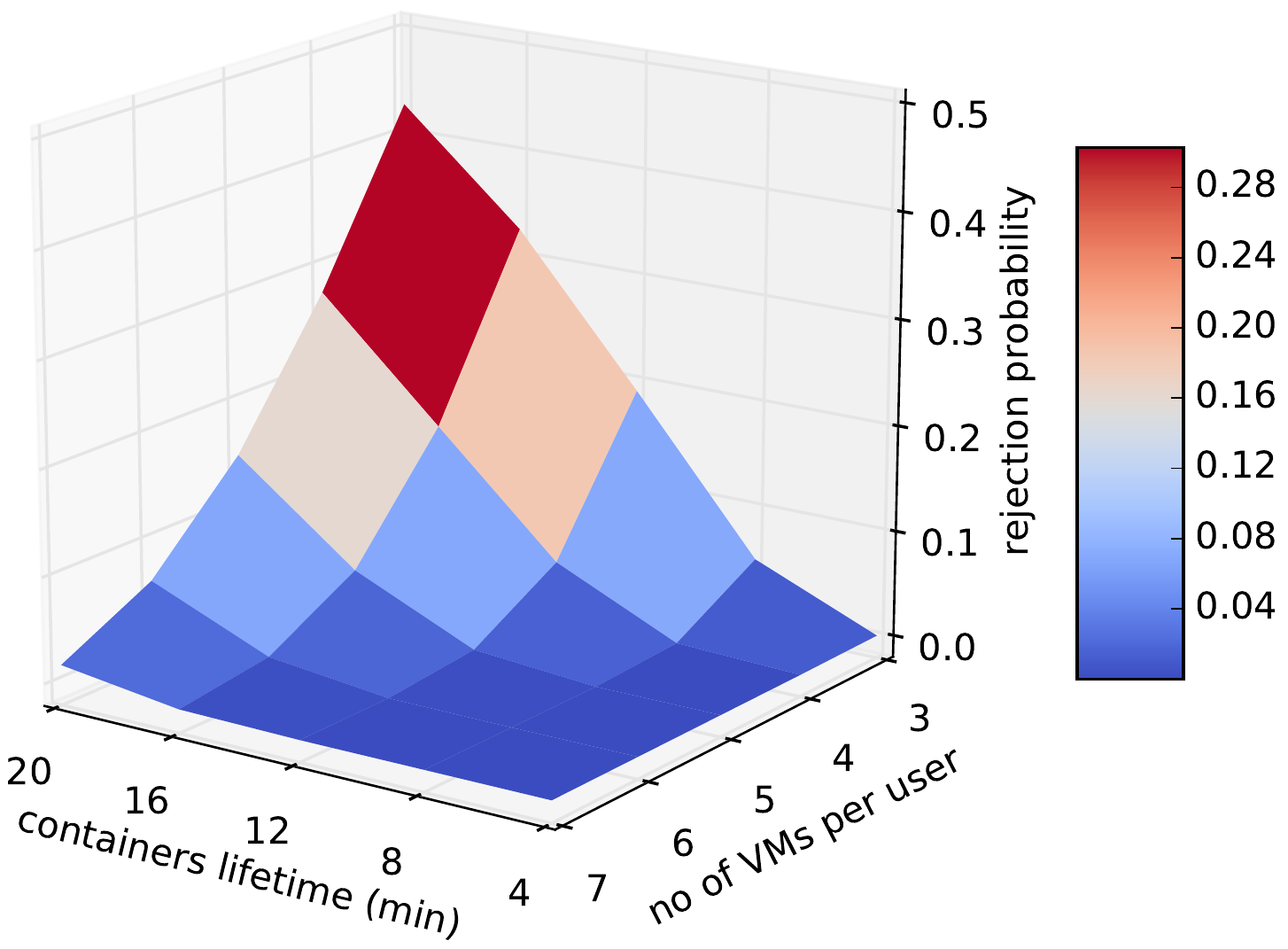}}
\subfigure[Request rejection probability in macro infrastructure.]
{\label{fig:rp-macro-first}
\includegraphics[width = 0.45\textwidth]{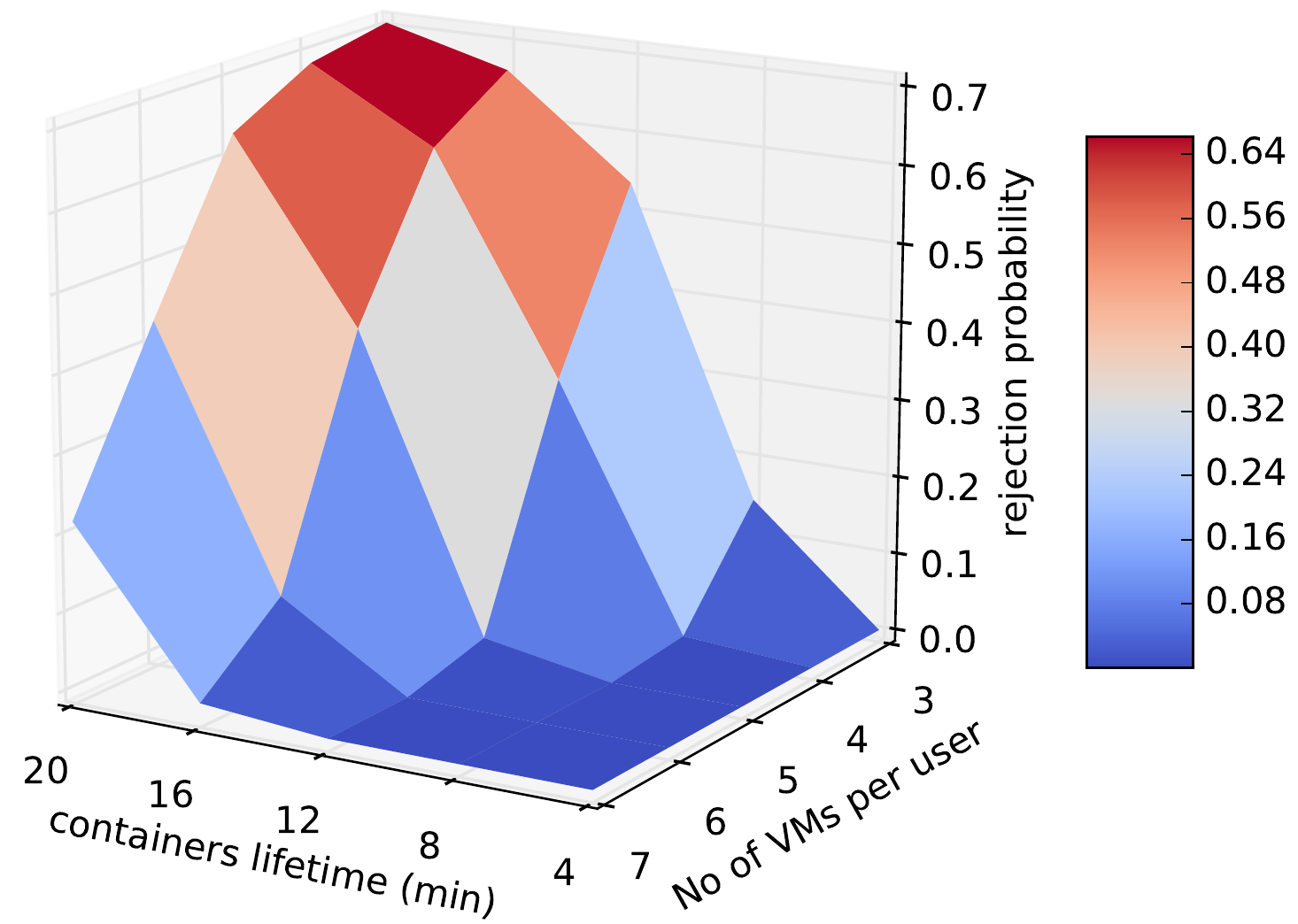}}
\end{center}
\caption{First scenario: rejection probability w.r.t containers' lifetime and the user's quota.}
\label{fig:rp-first}
\end{figure*}

\subsection{What-if Analysis and Capacity Planning}\label{what-if}
The presented performance model executed within a couple of seconds on a Macbook Pro with 16 GB of memory and 2.5 GHz quad-core Core-i7 CPU.
Thanks to this minimal cost and runtime associated with the analytical performance model, we leverage 
it to study interesting scenarios on a large scale with different configurations and parameter settings to shed some light on 
MSP provisioning performance. More specifically, in this section, we show how under different configurations and parameter 
settings, we obtain three important performance metrics, namely, rejection probability of requests, total response delay and
the probability of immediate service in both MSP and MSI. Note that due to space limitations, we only present a subset of results in 
this section. Table \ref{tab:si-range} presents the range of individual parameter values that we set for numerical analysis. 


\begin{table}[!ht]
\caption{Range of parameters for numerical experiment.}
\renewcommand{\arraystretch}{1.1}
\centering
\rowcolors{1}{white}{gray!25}
\begin{tabular}{|p{4cm}|cc|}
\hline
\multicolumn {3} {|c|} {\textbf{Microservice Platform (MSP)}} \\ 
\hline
\textbf{Parameter} & \textbf{Range} & \textbf{Unit} \\
\hline
No. of users in MSP & 20 & N/A\\
Arrival rate per user & $0.4\upto2$ & request/min \\
User's quota & $16\upto28$ & container \\
No. of containers on each VM & $4$ & N/A \\
Container provisioning time & $300\upto1500^*$ & millisecond \\
Container lifetime & $4\upto20$ & minute \\
Normal Host Group utilization &$40\%\upto80\%$ & N/A \\
Minimum Host Group size & 2 & VM \\
\hline
\multicolumn {3} {|c|} {\textbf{Macroservice Infrastructure (MSI)}} \\ 
\hline
\textbf{Parameter} & \textbf{Range} & \textbf{Unit} \\
\hline
Arrival rate & $ 60 $ & request/hour \\
VM lifetime & $1\upto5$ &  day \\
No. of PMs in the pool & 150 & PM \\
No. of VMs on each PM & 4 & VM \\
Mean look-up rate in the pool & 60 & search/min \\
VM provisioning time & $1\upto3^*$ & minute \\
Size of global queue & $100$ & requests  \\
\hline
\multicolumn{3}{|l|}{$^*$These values have been obtained from experiments.} \\
\hline
\end{tabular}
\label{tab:si-range}
\end{table}

\begin{figure*}[!ht]
\begin{center}
\subfigure[Total delay in microservice platform to obtain a container.]
{\label{fig:td-micro-first}
\includegraphics[width = 0.45\textwidth]{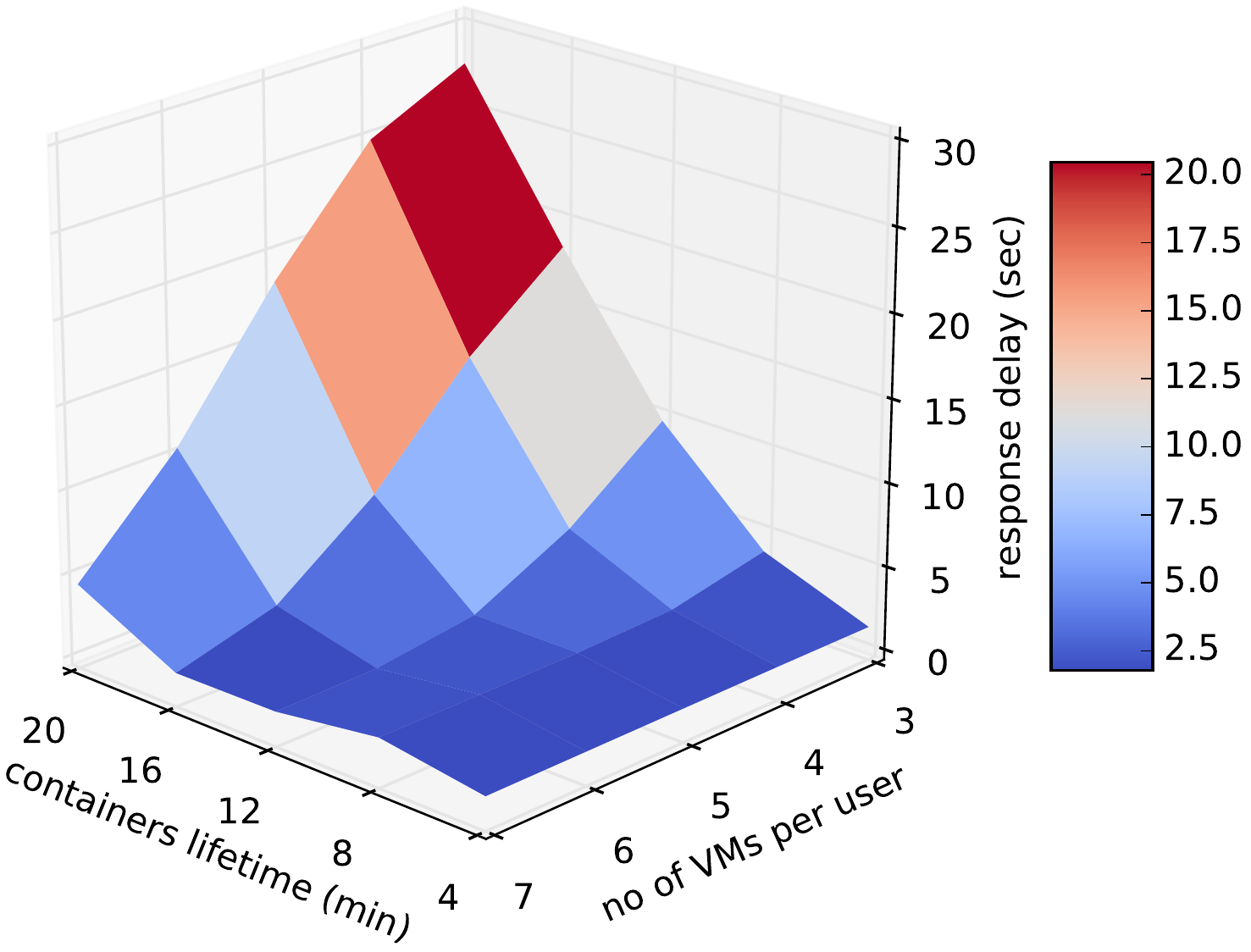}}
\subfigure[Total delay in macroservice infrastructure to obtain a VM.]
{\label{fig:td-macro-first}
\includegraphics[width = 0.45\textwidth]{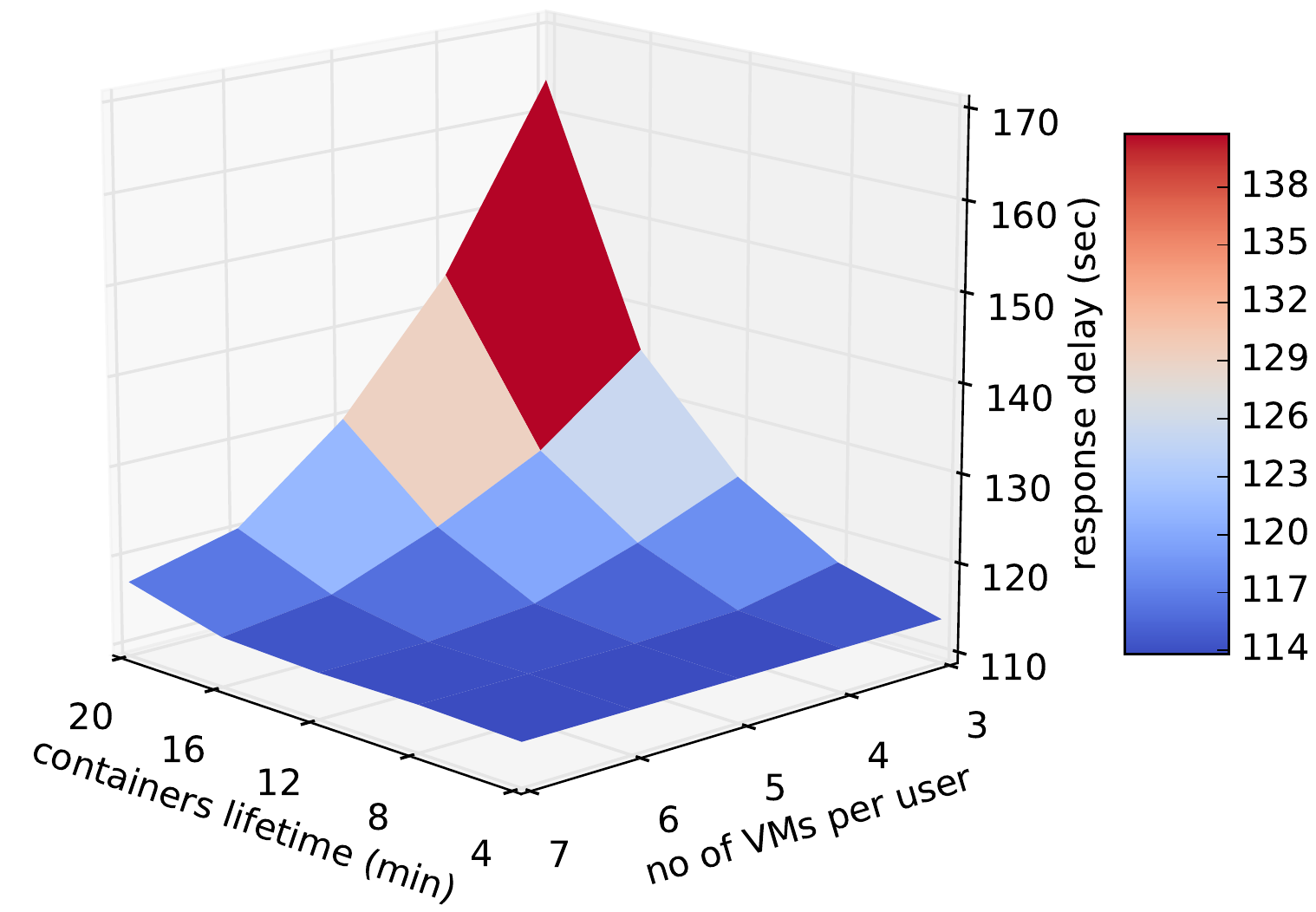}}
\end{center}
\caption{First scenario: total response delays w.r.t containers' lifetime and the user's quota;}
\label{fig:td-first}
\end{figure*}  

\subsubsection{First Scenario: Fixed Arrival Rate}

In the first scenario, we investigate the effects of container lifetime and the 
users' quota on rejection probability and total delay in both MSP and MSI. For the MSI, 
we set 2 days as the mean VM lifetime and assume 150 PMs in the servers' pool, each of which can 
run up to 4 VMs. In the first scenario, the results of which are shown in Fig. \ref{fig:rp-first}, as can be noticed,
both container lifetime and user quota have a significant impact on rejection probabilities. 
The impact of quota and lifetime is almost the same in MSP (Fig \ref{fig:rp-micro-first}) 
as well as MSI.  Also, in order to harness the rejection probability under 10\%, at least 4 VMs should be assigned to the 
user, and the container lifetime should be less than 12 minutes in the MSP. However, 
as can be seen from Fig. \ref{fig:rp-macro-first}, if the goal is to maintain less 
than 10\% rejection in MSI, the container lifetime should be less than 12 minutes, and 
at least 5 VMs should be assigned to the user's host group. From Fig. \ref{fig:rp-macro-first}, we can identify a very 
narrow area in which the backend cloud operates in a stable regime (i.e., the narrow dark blue rectangle) 
and will enter to unstable state with a little fluctuation either in containers' lifetimes or users' quota.

We also measure the delays in both micro and macro layers.
Note that the delay in microservice level is the aggregate waiting and processing 
time before having the containers ready for service; in the macro layer, it is all 
imposed delays on request before getting the VM up and ready to be used. Fig. \ref{fig:td-first} 
indicates that the trend of delays is in tune with rejection probabilities. From 
Fig. \ref{fig:td-micro-first}, it can be observed that if there is no need for a 
new VM, the request will be processed usually in less than a second, but if there 
is no VM that can accommodate the new container the delay might increase up to 27 seconds. 
Fig \ref{fig:td-macro-first} shows that, if the load is low, the macroservice 
infrastructure can provision a VM in 108 seconds on average (i.e., only processing time) but
as traffic intensity gets high it might take up to 170 seconds on average to get a VM. 

\begin{figure*}[!ht]
\begin{center}
\subfigure[Request rejection probability in microservice platform.]
{\label{fig:rp-micro-second}
\includegraphics[width = 0.47\textwidth]{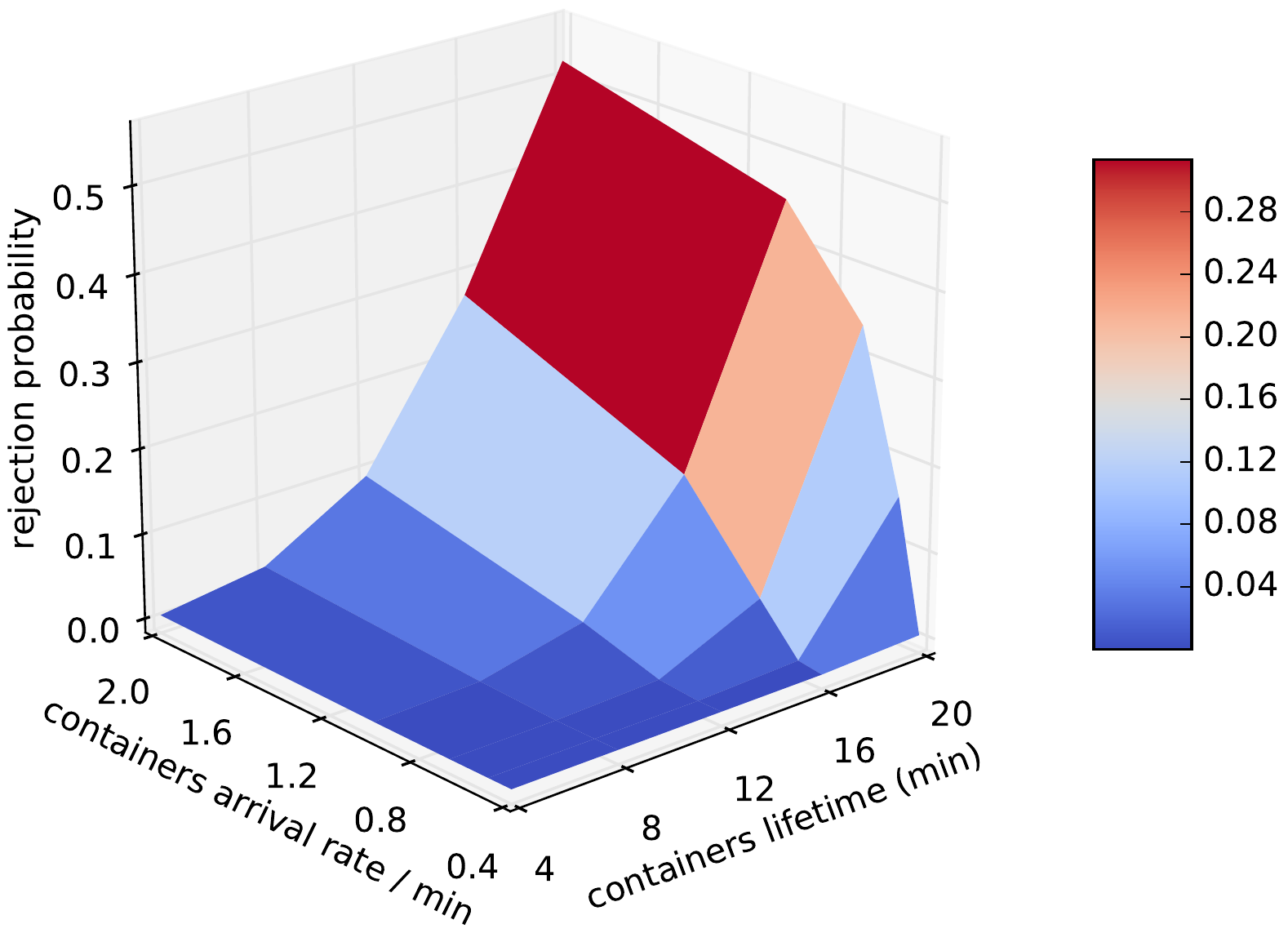}}
\subfigure[Total delay in microservice platform to acquire a container.]
{\label{fig:td-micro-second}
\includegraphics[width = 0.47\textwidth]{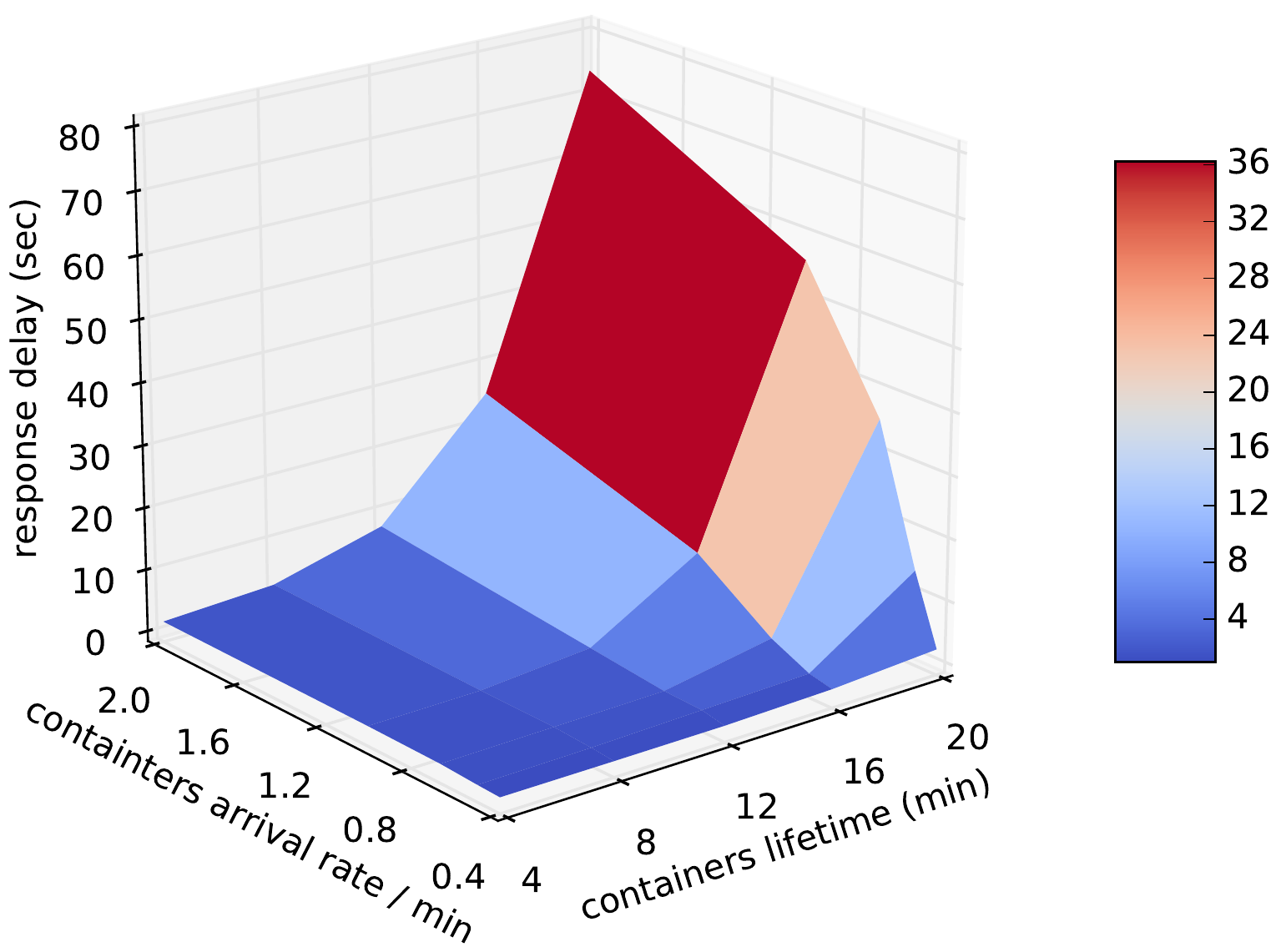}}
\end{center}
\caption{Second scenario: rejection probability and response delay w.r.t containers' lifetime and the arrival rate.}
\label{fig:rp-second}
\end{figure*}  

\subsubsection{Second Scenario: Fixed User Quota}

For the second scenario, we fixed the user quota at 16 containers and then studied the 
rejection probabilities and total delays by varying the container lifetime and arrival 
rate of requests for containers. 
This can help visualize the trade-off between lower provisioning time and higher
operational cost with having longer container lifetime.
For the macroservice infrastructure, we set 2 days as 
the mean VM lifetime and again assume 150 PMs in the pool. In this scenario,
we let each PM to run 4 VMs. From Fig. \ref{fig:rp-micro-second}, it can be noticed that
if the container lifetime increases linearly, the rejection probability increases exponentially 
while with increasing arrival rate, it increases sub-linearly. Therefore, if the user does 
not ask for more than 2 containers per minute and container lifetime stay bellow 12 minutes,
the rejection probability would be less than 10\%. 


From the diagram in Fig. \ref{fig:td-micro-second}, we can see that MSP imposes a 
long delay on requests in worst cases compared to the first experiment 
(i.e., Fig. \ref{fig:td-micro-first}). For example, in the microservice platform, when containers' lifetime 
and arrival rate are high (i.e., 20 minutes and 2 containers per minute respectively), 
the delay may bump up to 80 seconds. One potential solution for addressing this long delay is to permit MSP 
to ask for more than one VM at a time (i.e., making batch requests). 
\begin{figure*}[!t]
\begin{center}
\subfigure[First scenario.]
{\label{fig:imm-micro-first}
\includegraphics[width = 0.45\textwidth]{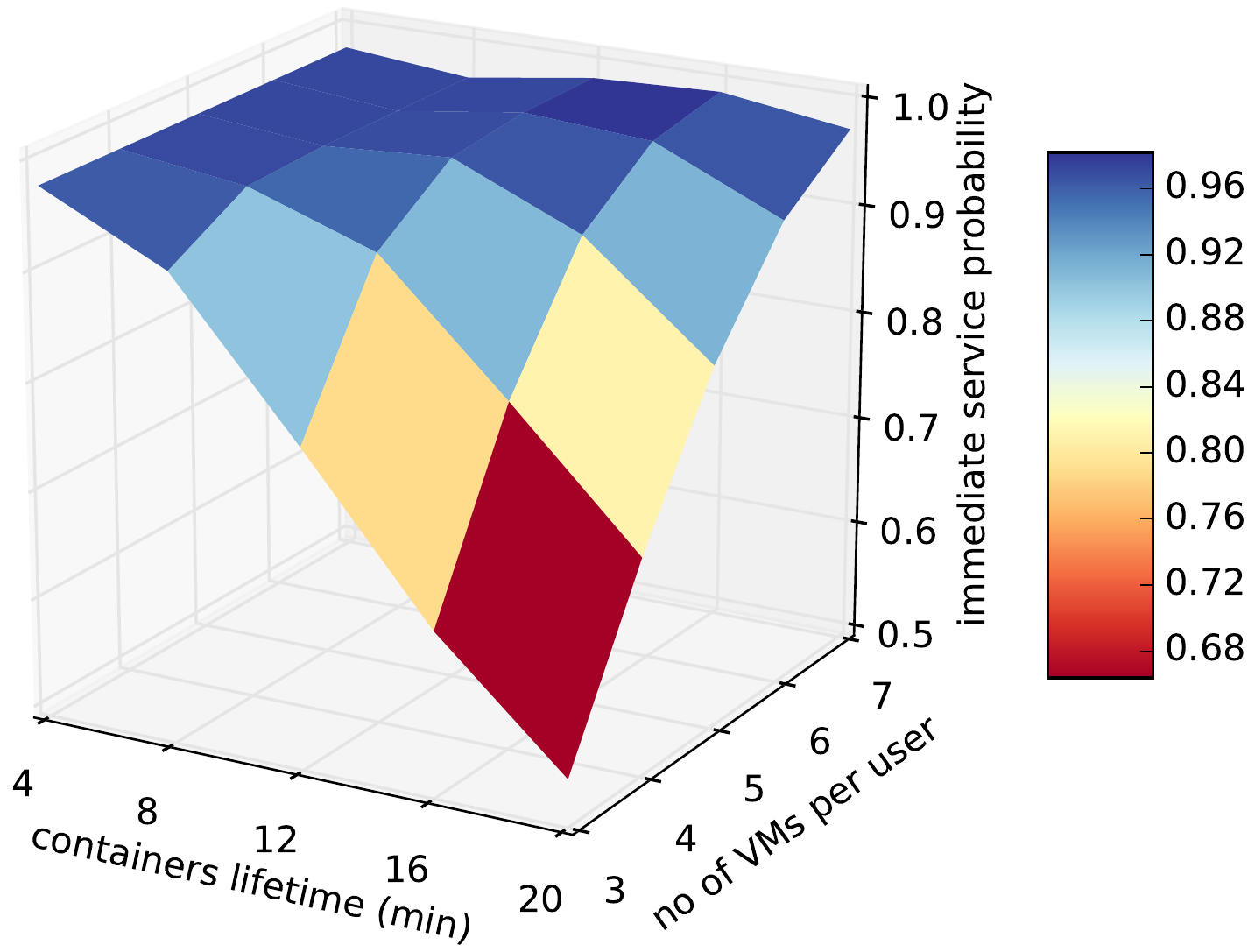}}
\hspace{0.5cm}
\subfigure[Second scenario.]
{\label{fig:imm-micro-second}
\includegraphics[width = 0.49\textwidth]{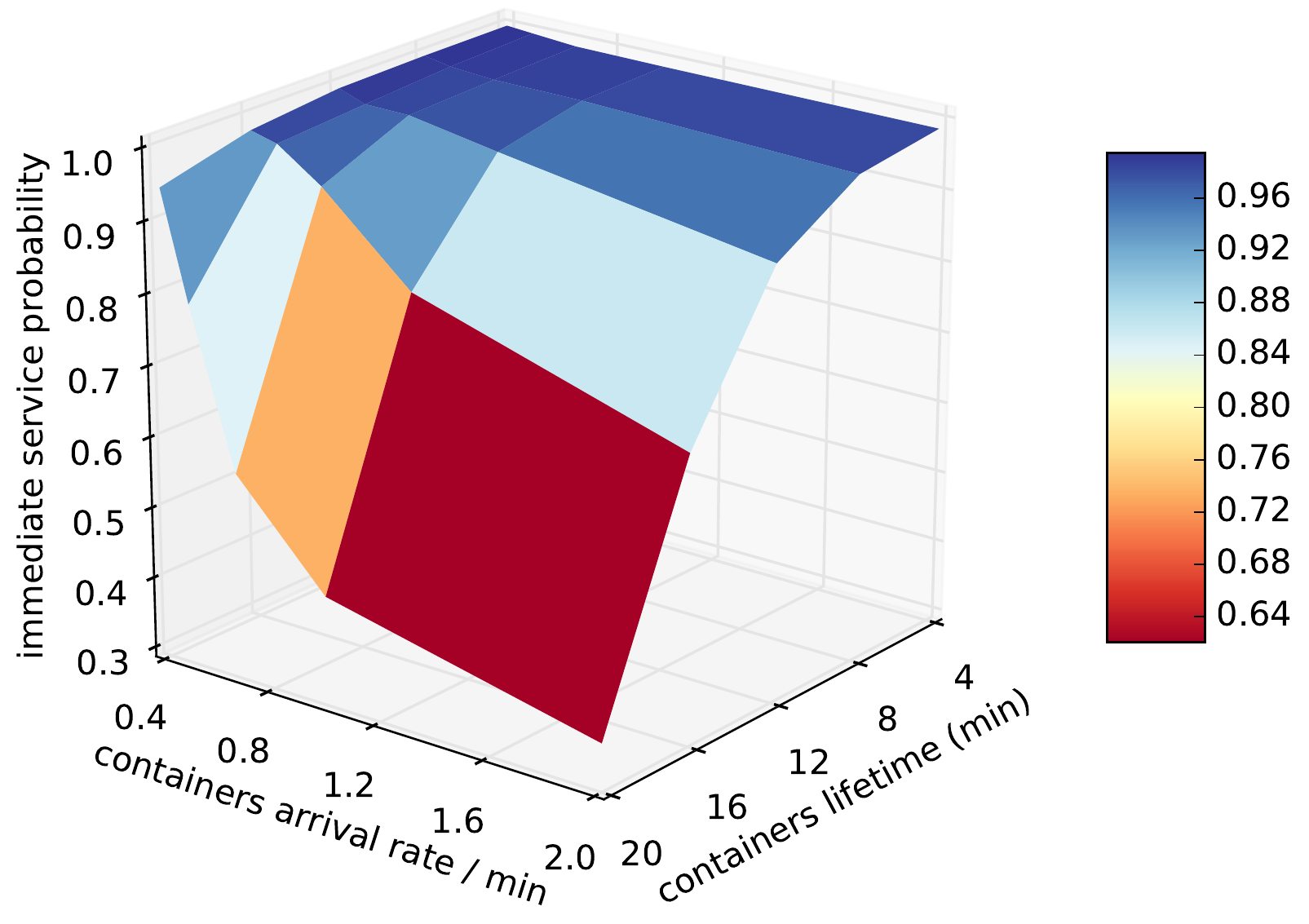}}
\end{center}
\caption{Probability of immediate service; probability by which requests get into service without any queuing.}
\label{fig:imm}
\end{figure*}  
We also characterize the probability of immediate service for both scenarios at both MSP and MSI. This metric 
shows how many of requests get into the service without any delay (i.e., queuing delay). As can be seen
in Fig. \ref{fig:imm}, the trend of immediate services probabilities are inverse to the trend of their corresponding rejection 
probabilities (i.e., compare Fig. \ref{fig:imm-micro-first} with \ref{fig:rp-micro-first} and Fig. \ref{fig:imm-micro-second} 
with \ref{fig:rp-micro-second}). However, it should be noted that these two probabilities are not truly complementing as there
are some requests that neither get rejected, nor immediate service rather get into services after some queuing.
Note that in the second scenario, we don't present the results for MSI due to the page limit. 

\section{Related Work} \label{rw}
Performance analysis of cloud computing has attracted considerable research attention, 
although most of the works considered hypervisor-based virtualization in which VMs are 
the sole way of providing an isolated environment for the users \cite{rochman2017dynamic, raei2017modeling}. 
However, recently, container-based virtualization has been getting momentum due to its 
advantages over VMs for providing microservices.

Performance analysis of IaaS clouds has been investigated extensively under various 
configurations and use cases. In \cite{hamzeh-tpds, hamzeh-tpds-high, bruneo2014stochastic, 
vakilinia2015modeling}, monolithic analytical performance models have been proposed and 
evaluated. An analytical model based on Markov chains to predict the number of cloud instances 
or VMs needed to satisfy a given SLA performance requirement such as response time, throughput, 
or request loss probability has been proposed in \cite{salah2015analytical}. In 
\cite{hamzeh-tpds-fine, hamzeh-tpds-pool}, the authors proposed a general analytical model for 
end-to-end performance analysis of a cloud service. They illustrated their approach using 
IaaS cloud with three pools of servers: hot, warm and cold, using service availability and 
provisioning response delays as the key QoS metrics. The proposed approach reduces the complexity of 
performance analysis of cloud centers by dividing the overall model into sub-models and 
then obtaining the overall solution by iteration over individual sub-model solutions. 

The authors in~\cite{gribaudo2018performance2} used a Fluid Petri Net model to
analyze the characteristics of a microservice platform. Research conducted in
\cite{gribaudo2017performance} analyzes large-scale microservice platforms
using simulated traces.
\nm{
Our work in this paper is based on \cite{khazaei2016efficiency}.
However, our previous work only models the microservice layer and treat the back-end
IaaS as a black box by assuming a stable performance by the back-end cloud
throughout time. But, our experiments with various cloud infrastructures revealed
that the throughput and efficiency of the back-end cloud infrastructure vary
event during short periods of time. 
Thus, in this work, we analyze all three levels of PMs, VMs, and container
with three sub-models which renders this work as an accurate model for
microservice platforms.
}

Performance analysis of cloud services considering containers as a virtualization option 
is in its infancy. Much of the works have been focused on the comparison between the implementation 
of various applications deployed either as VMs or containers. In \cite{joy2015performance}, 
authors made a performance comparison, including a front end application server hosting Joomla 
PHP application and backend server hosting PostgreSQL database for storing and retrieving 
data for the Joomla application. They showed that containers have outperformed 
VMs in terms of performance and scalability. Container deployment process 5x more
requests compared to VM deployment and also containers outperformed VMs by 22x in terms of 
scalability. This work shows promising performance when using containers instead of VMs for 
service delivery to end-users. 


Other works have focused on different aspects of microservice architecture.
The study performed in \cite{villamizar2016infrastructure} performs a cost 
comparison between running the same service in the cloud using
monolithic, microservice, and serverless architectures. The serverless
architecture in this study uses the microservices developed, but managed and
operated by the cloud provider. They found that microservices are a very
cost-effective way of deploying workloads to the cloud.
Heorhiadi et al.~\cite{heorhiadi2016gremlin} performed systematic resilience
testing of microservices. Inagaki et al.~\cite{inagaki2016container} studied
the scalability of different container management operations for docker and
identify scalability bottlenecks bottlenecks for this platform. Ueda et al.~\cite{ueda2016workload} performed a study to understand the characteristics
of different microservice workloads as means to design an infrastructure that
is optimized for microservices.

The authors in \cite{docker-perf-vsphere,felter2015updated} performed a more comprehensive study 
on performance evaluation of containers under different deployments. They used various benchmarks to study 
the performance of native deployment, VM deployment, native Docker and VM Docker. In native 
deployment, the application is installed in a native OS; in VM deployment, the application 
is installed in a vSphere VM; in native Docker, the application is installed in a container 
that is being run on a native OS; and finally, a Docker container including the application
is deployed on a vSphere VM that itself is deployed on a native OS. Redis has been used as the 
backend datastore in this experiment. All 
in all, they showed that in addition to the well-known security, isolation, and manageability 
advantages of virtualization, running an application in a Docker container in a vSphere VM 
adds very little performance overhead compared to running the application in a Docker container 
on a native OS. Furthermore, they found that a container in a VM delivers near-native performance
for Redis and most of the micro-benchmark tests.

These studies reveal a promising future for using both virtualization techniques in order to 
deliver secure, scalable and high performant services to the end-user 
\cite{gupta2015comparison, villamizar2015evaluating}. The recent popularity of microservice 
platforms such as Docker Tutum \cite{tutum}, Nirmata \cite{nirmata} and Google Kubernetes 
\cite{kubernetes} are attributed to such advantages mentioned above. However, to the best
of our knowledge, there is no comprehensive performance model that incorporates the details 
of the microservice platform and macroservice infrastructure. In this work, we studied the 
performance of PaaS and IaaS, collaborating with each other to leverage both virtualization 
techniques for providing fine-grained, secure, scalable and performant services.

\section{Conclusions} \label{con}
In this paper, we presented a performance model suitable for analyzing the provisioning quality 
of microservice platforms while incorporating the details of servicing in the backend IaaS cloud using 
interacting stochastic models. We have developed a comprehensive analytical model that captures 
important aspects of the system including microservices management, resource assigning process, and 
virtual machine provisioning. The performance model can assist cloud micro/macro service 
providers to maintain their SLA in a systematic manner. In other words, the proposed and evaluated
performance model in this paper provides a systematic approach to study the elasticity of  
microservice platforms by evaluating the provisioning performance at both the microservice platform and 
the back-end cloud.

We have also implemented a microservice platform from scratch to estimate related parameters to be used for
calibration of the analytical model. After introducing the measurements provided by the real implementation into
the performance model, we carried out an extensive numerical analysis to study the effects of various parameters 
such as the request arrival rate for containers, container lifetime, VM lifetime, virtualization 
degree, and the size of the users' application on the request rejection probability, probability of immediate service 
and response time. We showed that using the performance model, we can characterize the behaviour 
of the system at scale for given configurations and therefore facilitate the capacity planning and SLA 
analysis by both micro and macro service providers. 

\section*{Acknowledgments} 
We would like to thank Dr. Murray Woodside for his valuable technical comments and inputs. 
This research was supported by the Natural Sciences and Engineering Council of Canada (NSERC).




%

\vspace{-1.2cm}

\begin{IEEEbiography}[{\includegraphics[width=1in,height=1.25in,clip,keepaspectratio]
{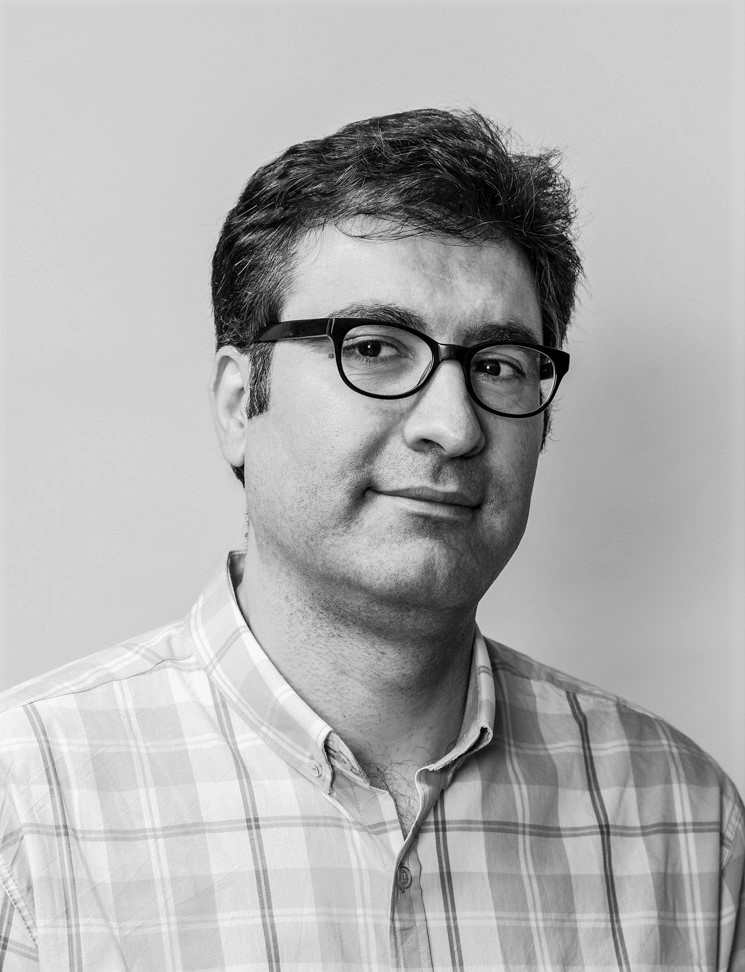}}]{Hamzeh Khazaei, PhD.} is an assistant professor in the Department of Electrical Engineering and Computer Science at York University. Previously he was an assistant professor at the University of Alberta, a research associate at the University of Toronto and a research scientist at IBM, respectively. He received his PhD degree in Computer Science from the University of Manitoba, where he extended queuing theory and stochastic processes to accurately model the performance and availability of cloud computing systems. His research interests include performance modelling, cloud computing and engineering distributed systems. 
\end{IEEEbiography}
\vspace{-1.2cm}

\begin{IEEEbiography}[{\includegraphics[width=1in,height=1.25in,clip,keepaspectratio]{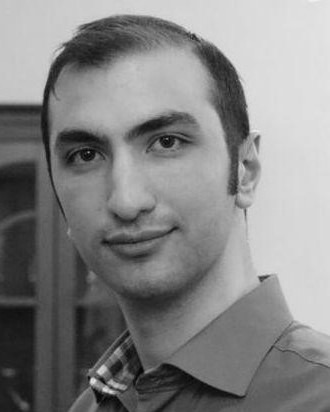}}]{Nima Mahmoudi}
received two BS degrees in Electronics and Telecommunications
and his MS degree in Digital Electronics from
Amirkabir University of Technology, Iran in 2014, 2016, and 2017 respectively.
He is currently working towards the PhD degree in software engineering and intelligent systems at the University of Alberta.
He is a Research Assistant at the University of Alberta and a visiting Research Assistant in the Performant and Available Computing Systems (PACS) lab at York University.
His research interests include serverless computing, cloud computing, performance modelling, applied machine learning, and distributed systems. He is a graduate student member of the IEEE.
\end{IEEEbiography}
\vspace{-1.2cm}

\begin{IEEEbiography}
[{\includegraphics[width=1in,height=1.25in,clip,keepaspectratio]{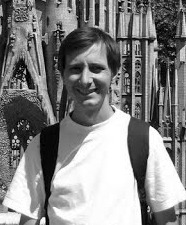}}]{Cornel Barna, PhD.}
is a Professor at Seneca Collage, School of Information and Communications Technology. Prior to joining Seneca 
Collage, he was a postdoctoral fellow at York 
University where he obtained his PhD in computer science as well. He obtained his MSc and BSc in computer science
from Cuza University, Romania. His main research area is adaptive algorithms,
web applications in cloud, heuristics \& metaheuristics, and management of containerized software.
\end{IEEEbiography}

\vspace{-1.2cm}
\begin{IEEEbiography}
[{\includegraphics[width=1in,height=1.25in,clip,keepaspectratio]{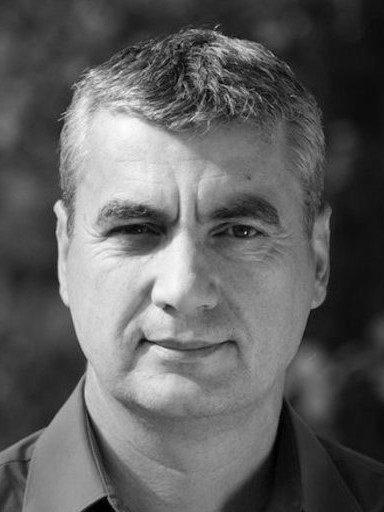}}]{Marin Litoiu, PhD, Peng.}
is a Professor in the Department of Electrical Engineering and Computer Science and in the 
School of Information Technology, York University. He leads the Adaptive Software Research Lab and focuses on 
making large software systems more versatile, resilient, energy-efficient, self-healing and self-optimizing. His 
research won many awards including the IBM Canada CAS Research Project of the Year Award and the IBM 
CAS Faculty Fellow of the Year Award for his “impact on IBM people, processes and technology.”
\end{IEEEbiography}

\end{document}